\documentclass[letterpaper,journal]{IEEEtran}
\usepackage{amsmath,amsfonts}
\usepackage{algorithmic}
\usepackage{algorithm}
\usepackage{array}
\usepackage[caption=false,font=normalsize,labelfont=sf,textfont=sf]{subfig}
\usepackage{textcomp}
\usepackage{stfloats}
\usepackage{url}
\usepackage{verbatim}
\usepackage{graphicx}
\usepackage{cite}
\usepackage{tabularx} 
\usepackage[most]{tcolorbox} 
\usepackage{multirow}
\usepackage{makecell} 
\usepackage{xspace} 
\usepackage{etoolbox}
\usepackage{threeparttable}
\usepackage{ragged2e}

\begin{document}

\title{An Empirical Study on Virtual Reality Software Security Weaknesses}


\author{Yifan Xu, Jinfu Chen, Zhenyu Qi, Huashan Chen, Junyi Wang, Pengfei Hu, Feng Liu, Sen He
\thanks{Yifan Xu, Huashan Chen, and Feng Liu are with the Institute of Information Engineering, Chinese Academy of Sciences, and also with School of Cyber Security, University of Chinese Academy of Sciences, Beijing, China (e-mail: xuyifan@iie.ac.cn; chenhuashan@iie.ac.cn; liufeng@iie.ac.cn).}
\thanks{Jinfu Chen is with the School of Computer Science, Wuhan University, Wuhan, China (e-mail: jinfuchen@whu.edu.cn).}
\thanks{Zhenyu Qi and Sen He are with the Department of Electrical and Computer Engineering, University of Arizona, Tucson, USA (e-mail: qzydustin@arizona.edu; senhe@arizona.edu).}
\thanks{Junyi Wang and Pengfei Hu are with School of Computer Science and Technology, Shandong University, Qingdao, China (e-mail:junyiwang@sdu.edu.cn, phu@sdu.edu.cn).}

\thanks{Corresponding author: Huashan Chen.}
}

\markboth{IEEE TRANSACTIONS ON DEPENDABLE AND SECURE COMPUTING,~Vol.~*, No.~* July~2025}
{Shell \MakeLowercase{\textit{et al.}}: An Empirical Study on Virtual Reality Software Security Weaknesses}

\IEEEpubid{0000--0000/00\$00.00~\copyright~2025 IEEE}

\maketitle

\begin{abstract}
Virtual Reality (VR) has emerged as a transformative technology across industries, yet its security weaknesses, {including vulnerabilities,} are underinvestigated. 
This study investigates 334 VR projects hosted on GitHub, examining 1,681 software security weaknesses to understand: {\em what} types of weaknesses are prevalent in VR software; {\em when} and {\em how} weaknesses are introduced; {\em how long} they have survived; and {\em how} they have been removed. Due to the limited availability of VR software security weaknesses in public databases (e.g., the National Vulnerability Database or NVD), we prepare the {first systematic} dataset of VR software security weaknesses by introducing a novel framework to collect such weaknesses from GitHub commit data.
Our empirical study on the dataset leads to useful insights, including: (i) VR weaknesses are heavily skewed toward user interface weaknesses, followed by resource-related weaknesses; (ii) VR development tools pose higher security risks than VR applications;
(iii) VR security weaknesses are often introduced at the VR software birth time.
\end{abstract}


\vspace{-0.3cm}
\section{Introduction}

{

\IEEEPARstart{V}{irtual} Reality (VR) technology has undergone significant advancements in recent years, with its applications cutting across many sectors such as gaming, education, healthcare, and remote collaboration \cite{GobbettiVirtual, Smutny2023Learning, Creed2024Inclusive, Javaid2020Virtual, Akbulut2018effectiveness, Martin-Gutierrez2017Virtual, Weiss1998Virtual, Pujiono2024Augmented, Berg2017Industry}. 
While the VR hardware market continues to grow,
VR software is increasingly gaining momentum as evidenced by the fact that thousands of applications are becoming available on major app distribution platforms (e.g., Google Play, Apple Store, Oculus), with approximately 200 million downloads worldwide~\cite{VirtualR14:online}. 

Despite the rapid growth of VR hardware and software, VR security has not been paid the due amount of attention despite its potential damages 
\cite{Dastgerdy2024Virtual, Vondracek2023Rise, Silva2023Survey}. For instance, there is little understanding of the unique nature of VR software when compared to traditional software
\cite{Elliott2015Virtual}; understanding their difference is important because existing software security mechanisms may not be adequate to harden VR software. 
This is possible because VR software has two unique features: (i) the real-time data processing and immersive user interactions of VR may incur new kinds of attack surfaces; and 
(ii) the intricate integration of hardware and software in VR environments may exacerbate VR security.

Previous studies on VR weaknesses mainly focus on specific types of software weaknesses \cite{Guo2025Empirical,Rodriguez2017Empirical,Rzig2023Virtual},
weaknesses associated with physical devices \cite{Dastgerdy2024Virtual, chandrashekar2023design, siddiqi2023secure},
or weaknesses associated with specific development platforms, frameworks, and applications \cite{Vondracek2023Rise, qamar2023systematic, nnamonu2023metaverse}. 
However, there is no study on the landscape of VR software security weaknesses.
For instance, 
there is no systematic study on {\em what} types of security weaknesses are prevalent in VR software, {\em when} and {\em how} those security weaknesses are introduced, {\em how long} they have survived, and {\em how} some of them may have been removed. This motivates the present study.

\IEEEpubidadjcol 

This study aims to fill the gap through an 
empirical analysis of VR software security weaknesses, referred to as {\em VR weaknesses} thereafter.
However, existing databases, such as the NVD~\cite{NVD} and VulDB~\cite{VulDB}, contain few VR weaknesses because the community has not made the effort to systematically identify them for incorporation into these databases. To address this problem, we introduce a novel framework for identifying VR weaknesses from the VR software on GitHub by analyzing repository metadata, commit messages, and file modifications.
To enhance the quality of the resulting VR weaknesses dataset, we use multiple semantic models for cross-validation and reducing false-positives. This leads to a dataset of 1,681 VR weaknesses associated with 334 open-source VR projects. 

Based on the dataset, we formulate a set of research questions (RQs) on the lifecycle, distribution, detection, and fixing of VR weaknesses. Our empirical analysis leads to interesting findings, such as: (i) VR weaknesses are heavily skewed toward user interface weaknesses, followed by resource-related weaknesses; (ii) VR development tools pose higher security risks than VR applications;
(iii) VR weaknesses are often introduced at the VR software birth time;
(iv) VR weaknesses are typically short-lived and resolved via localized, additive changes affecting only a few files; (v) Traditional static and dynamic analysis methods may be less effective at detecting VR weaknesses due to the unique characteristics of VR software.

To sum up, this paper makes the following contributions:
\begin{itemize}
    \item We introduce a novel framework of identifying security weaknesses from GitHub, leading to the first systematic VR weaknesses dataset. Both the framework and the dataset would be of independent value.

    \item We conduct an empirical analysis on the dataset via nine RQs to explore the characteristics of VR weaknesses.

    \item We draw a number of insights into VR weaknesses and discuss their implications.

\end{itemize}

\textbf{Paper Organization.} Section \ref{sec:background} discusses preliminary knowledge on VR.
Section \ref{sec:method} describes our framework, including the method for identifying security weaknesses from GitHub.
Section \ref{sec:dataset_collection} introduces our dataset collection process and shows the specific data.
Section \ref{sec:evaluation} presents our empirical analysis.
Section \ref{sec:discussion} discusses recommendations, limitations, and applicability of the study. 
Section \ref{sec:related_work} reviews related prior studies. Section \ref{sec:conclusion} concludes the paper.

\vspace{-0.3cm}
\section{Background Knowledge on VR}
\label{sec:background}

\subsection{VR Architecture and Workflow}
\label{sec:vr}
The VR technology enables a first-person and immersive digital experience by integrating hardware and software components to simulate interactive, three-dimensional environments. To meet the unique challenges of real-time rendering, complex user interactions, and high resource dependencies, VR systems require advanced system architectures and well-thought-out development planning to ensure their sustainability and adaptability in real-world applications. The framework of VR technology is illustrated in Figure~\ref{fig:vr_framework}.

\begin{figure}[!htbp]
    \centering
    \includegraphics[width=0.9\linewidth]{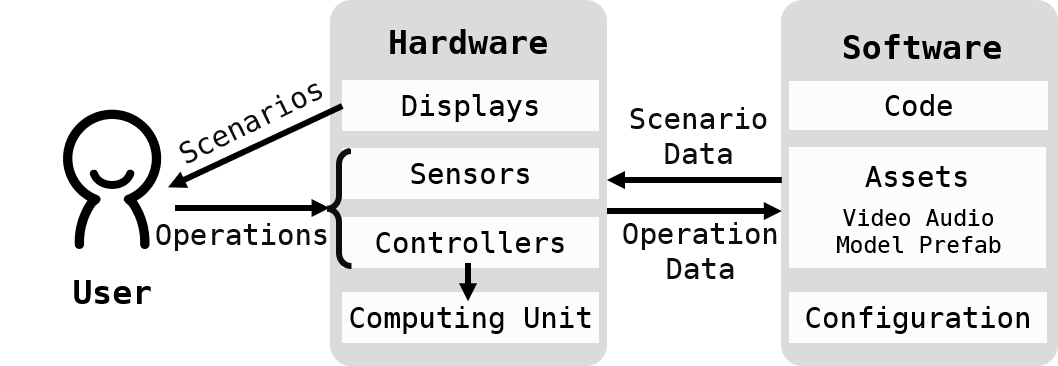}
    \caption{Architecture and workflow of VR technology.}
    \label{fig:vr_framework}
    \vspace{-0.2cm}
\end{figure}

\textbf{Architecture.}
The VR system architecture combines hardware and software elements that work together to produce an immersive user experience. The hardware includes displays (e.g., immersive screens, head-mounted displays), sensors (e.g., gyroscopes, accelerometers, external tracking cameras), controllers (e.g., motion-tracking devices, haptic feedback systems), and computing units (e.g., GPUs, personal computers, or standalone VR devices). On the software side, VR software is typically divided into tool-based and application-based categories. The software consists of core code files, generally written in C\#, which define the primary functionalities of the system, along with asset files (e.g, videos, audio, models, prefabs) and configuration files, which include both a global configuration file for the entire system and individual sub-configuration files for each asset.

\textbf{Workflow.}
In VR systems, user movements are continuously tracked, and sensory feedback is dynamically adjusted through an integrated loop involving hardware, software, and real-world interactions \cite{How-2025-03-17}. Sensors and controllers detect user movements, gestures, and other inputs, transmitting this data to the computing unit, where it is processed in real time and sent to the software for logical analysis. The software uses this real-time data to calculate the updated scenario and its associated information, which is then sent back to the computing unit. The computing unit then updates the displays with the new scenario. Based on this updated environment, the user proceeds with their next action, and the sensors and controllers continue capturing new data for the next cycle.

\vspace{-0.5cm}
\subsection{Unique Characteristics of VR Software}
\label{sec:difference}

VR software focuses on immersive experience and real-time interaction, making it different from traditional desktop applications in both architecture and runtime characteristics.

\textbf{Scene Oriented Structure.}
Desktop applications are organized around UI components and functional modules. However, VR software is structured around different virtual scenes. Scenes are independent of each other, and the switching of scenes is driven by users. This structure means that most code is bound to one scene, except for basic public code such as rendering. Each scene has its own interaction logic, and the code scripts in the scene run independently.

\textbf{Real-time Multi-Source Interaction.}
The input signals of desktop applications are mainly from the mouse and keyboard. However, VR software has many sensors with rich interaction signals, such as position, motion, perspective, and sound. Similar to scene-oriented structure, these interaction signals are bound with assets in the scene. The real-time frequent interaction of the user can generate signals constantly, meaning that different assets are constantly invoked and released.

\textbf{High Performance Requirements.}
In the desktop application, minor delays or resource inefficiencies rarely disrupt the overall functionality, especially on modern computers with high-performance CPUs and GPUs. However, VR software requires more stringent real-time performance. To ensure real-time rendering and data processing under limited hardware conditions, VR developers often use multithreading to optimize GPU usage and minimize memory overhead.

\vspace{-0.3cm}
\section{The Framework}
\label{sec:method}

Figure~\ref{fig:overall} illustrates our analytical framework, comprising three core modules: {(i) {\em attributes definition}, (ii) {\em dataset construction}, and (iii) {\em RQs (Research Questions) definition}.} These components are detailed below.

{\bf Terminology}. In this paper, we use the term {\em weakness} rather than {\em vulnerability} because we align our study to the Common Weakness Enumeration (CWE) \cite{CWE} framework. This is reasonable because the CWE framework offers the concept of weakness types rather than speaking of individual vulnerabilities. Nevertheless, CWE includes vulnerabilities. 
We use the terms {\em VR open-source software} and {\em VR software} interchangeably as the corpus collected from GitHub, which includes {\em VR applications} and {\em VR development tools}.

\begin{figure*}[t]
    \centering    \includegraphics[width=\linewidth]{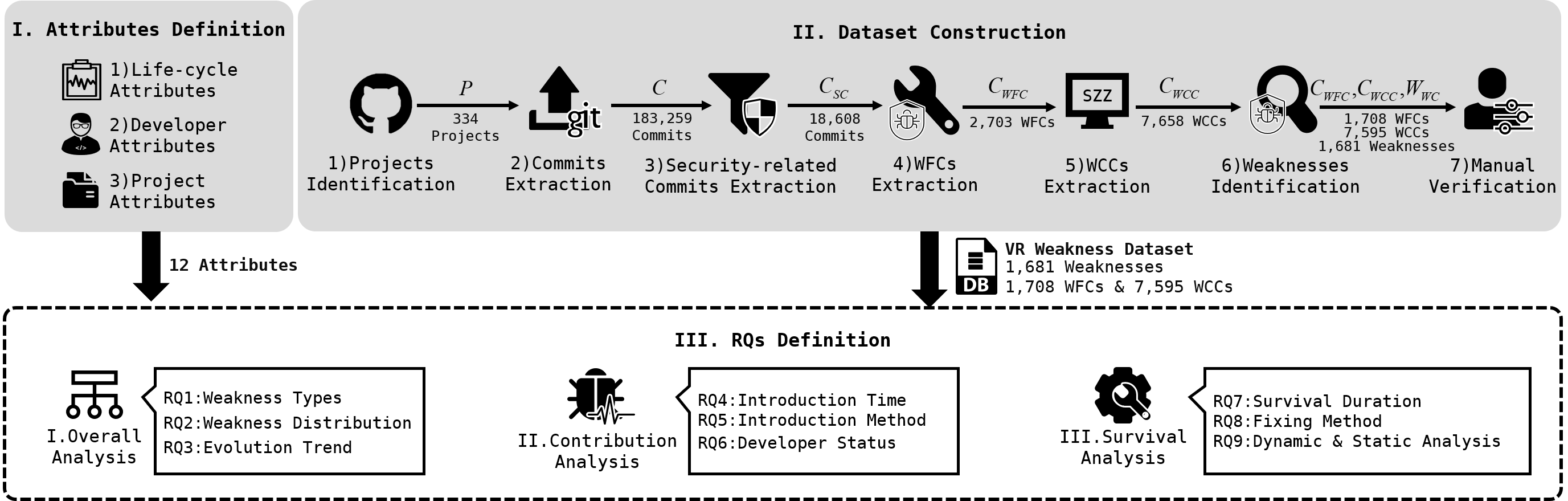}
    \caption{Framework overview.}
    \label{fig:overall}
    \vspace{-0.4cm}
\end{figure*}

\vspace{-0.1cm}
\subsection{Attributes Definition}
\label{sec:attributes}

To enable a thorough quantitative analysis, we define three groups of attributes: 8 attributes characterizing VR weakness lifecycle, 2 attributes reflecting developer status, and 2 attributes delineating project status. 

\subsubsection{VR Weakness Lifecycle Attributes}
\label{sec:lifecycle_definations}
\ 
\newline
\indent 
As depicted in Figure \ref{fig:vulnerability_lifecycle}, the lifecycle of a VR weakness typically includes four specific time points, forming several critical windows or phases.

{\bf Time Points}. To trace the milestones from the introduction to the resolution of a VR weakness, as informed by prior studies \cite{meneely2013patch, Iannone2023Secret}, we adopt the term WCCs (weakness-contributing commits) to describe the commits in the version control repository that contribute to the introduction of a post-release VR weakness. Additionally, we propose the term WFCs (weakness-fixing commits) to denote the commits that resolve an identified VR weakness.

\begin{itemize}
\item \textbf{Origin Point (\(T_0\))} denotes the time at which a VR weakness is first introduced into the project, corresponding to the first WCC where the initial flawed or harmful code is added to the repository. This point marks the beginning of the weakness lifecycle, establishing a reference for tracking its subsequent impact and persistence within the project. 

\item \textbf{Turning Point (\(T_1\))} denotes the time associated with the last WCC, signifying when the VR weakness is fully integrated into the repository and starts influencing the project \cite{Iannone2023Secret}.

\item \textbf{Expose Point (\(T_2\))} denotes the time linked to the first WFC of the VR weakness, reflecting the initiation of corrective measures and highlighting when efforts to mitigate the security risk commence.

\item \textbf{Resolution Point (\(T_3\))} denotes the time corresponding to the commit where the VR weakness is finally fixed, highlighting the point when the issue is resolved in the repository. 
This point marks the end of the weakness lifecycle, showing that corrective actions have been implemented to mitigate the security risk. It is important to note that for weaknesses fixed only once, 
\(T_2\) and \(T_3\) might coincide in time. 
\end{itemize}

{\bf Windows}. The four windows measure the time intervals between these critical points, capturing the duration of each phase in the lifecycle. 

\begin{itemize}
    \item \textbf{Insertion Window (\(t_{0,1}\))} represents the time interval between origin point \(T_0\) and turning point \(T_1\) \cite{Iannone2023Secret}, indicating the phase during which the VR weakness is introduced within the repository. This phase provides insights into the initial period of VR weakness exposure, tracking how vulnerabilities may grow or change before fixing. Formally, the insertion window is defined as \(t_{0,1}=T_1-T_0\).
    
    \item \textbf{Latency Window (\(t_{1,2}\))} represents the time interval between the turning point \(T_1\) and the expose point \(T_2\), reflecting the phase during which the VR weakness remains in the repository before any corrective action is initiated. This window measures the latency in identifying and starting to address the VR weakness. Formally, the detection window is defined as \(t_{1,2}=T_2-T_1\).
    
    \item \textbf{Fixing Window (\(t_{2,3}\))} represents the time interval between the exposure point \(T_2\) and the resolution point \(T_3\), quantifying the time taken to fully address the VR weakness after its initial fix. This metric sheds light on the complexity of resolving the VR weakness. Formally, the detection window is defined as \(t_{2,3}=T_3-T_2\).
    
    \item \textbf{Lifetime (\(t_{1,3}\))} represents the time interval between the turning point \(T_1\) and the resolution point \(T_3\), measuring the duration the VR weakness persists within the repository before being completely fixed. 
    These metrics reflect the persistence of the VR weakness within the project.
    Formally, the lifetime is defined as \(t_{1,3}=T_3-T_1\). 
    Notably, when \(T_2\) and \(T_3\) occur simultaneously, the latency window \(t_{1,2}\) and the lifetime \(t_{1,3}\) align.
\end{itemize}

\begin{figure}[!htbp]
    \centering
    \vspace{-0.3cm}
    \includegraphics[width=0.9\linewidth]{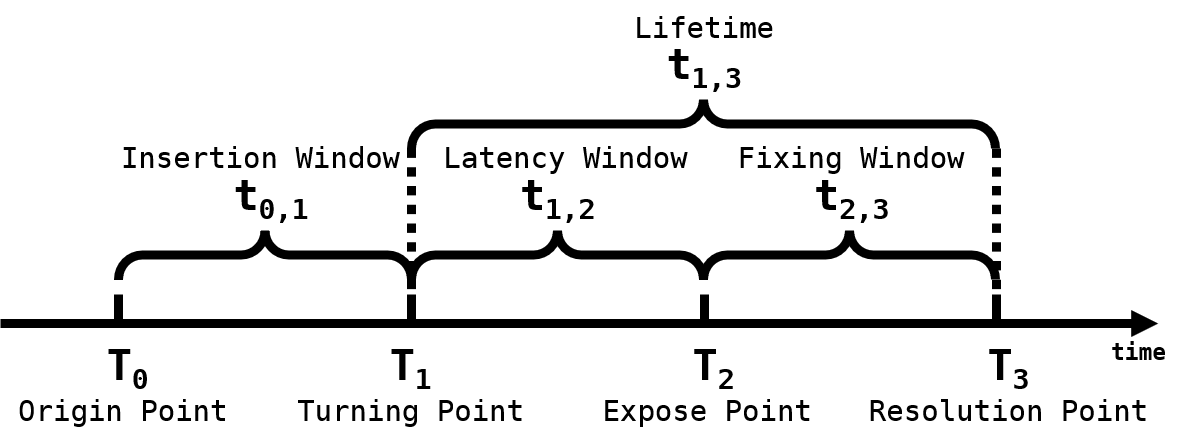}
    \caption{The lifecycle of VR weakness.}
    \label{fig:vulnerability_lifecycle}
    \vspace{-0.3cm}
\end{figure}

\subsubsection{Developer Attributes}
\label{sec:develop_definations}
\ 
\newline
\indent 
To explore the relationship between developers and VR weakness introduction, we adopt two developer-related attributes that are widely referenced in prior studies.

\begin{itemize}
\item \textbf{Workload (\(WL\))} refers to the proportion of a submitting author's contributions relative to the total contributions made by all authors in the preceding 30 days \cite{Iannone2023Secret}, measured either by the ratio of the number of commits—termed \textit{commit workload} (\(WL_{commit}\))—or by the ratio of the code change size—termed \textit{code chunk workload} (\(WL_{code}\)).
In line with prior studies, \(WL_{commit}\) is assigned as {\em `low'} ($<0.25$), {\em `medium'} ($0.25\sim0.75$), or {\em `high'} ($>0.75$), with a similar mapping for \(WL_{code}\).  

\item \textbf{Experience (\(Exp\))}, expressed as a proportion, measures the developer's contribution period from his first recorded activity in the project to the completion of a given WCC, relative to the total contribution time of all authors \cite{Iannone2023Secret}.
It can classify developers as {\em `newcomer'} ($<0.25$), {\em `medium'} ($0.25\sim0.75$), or {\em `expert'} ($>0.75$). 
\end{itemize}

\subsubsection{Project Attributes}
\label{sec:project_definations}
\ 
\newline
\indent 
We design two project-level attributes to describe the general status of the projects. 
\begin{itemize}
\item \textbf{Active Duration (\(D_{act}\)),} expressed in months, measures the period from the first commit to the last commit of the repositories, representing the full-time span during which the project is actively receiving contributions.

\item \textbf{Commit Frequency (\(F_{com}\))} refers to the total number of commits made within a specific period, reflecting the project's activity level since its creation.
The metric is formally defined as \(F_{com}=N_{com}/D_{act}\), where \(N_{com}\) represents the total number of commits.
\end{itemize}

\subsection{Dataset Construction}
\label{sec:dataset_construction}
{\color{black}In light of the lack of publicly documented VR weaknesses in well-established vulnerability databases such as NVD~\cite{NVD}, VulDB~\cite{VulDB}, and Exploit Database~\cite{ExploitDB}, we propose a novel method that identifies such weaknesses by exclusively mining and analyzing commit histories from open-source GitHub repositories, and compiles these weaknesses into a curated dataset for empirical analysis. The method unfolds in seven stages, outlined as follows.

\subsubsection{Projects Selection:}
The first stage focuses on selecting VR-related GitHub repositories for analysis. The selection process leverages two complementary sources: existing benchmark datasets reported in the literature and additional repositories retrieved through domain-specific keyword searches on GitHub. To ensure the inclusion of meaningful and actively maintained projects, we apply a filtering criterion that retains only those with a reasonable number of historical commits. The resulting set of qualified projects is denoted as $P$.

\subsubsection{Commits Extraction:}
The second stage entails extracting historical commits and their associated metadata from each repository in the project set $P$. To this end, we locally clone each GitHub-hosted project and retrieve its full commit history.
In parallel, we collect complementary repository metadata, including repository creation timestamps, primary programming languages, contributor identities, and popularity indicators (e.g., stars and forks). The resulting set of commits, denoted as $C$, is serialized into a structured \verb|json| format to support subsequent analysis.
for further analysis.

\subsubsection{Security-related Commits Extraction:}
The third stage is dedicated to identifying security-related commits (SCs), resulting in a subset of the full commit set $C$, denoted as $C_{SC}$, where $C_{SC} \subseteq C$. For this purpose, we employ a hybrid detection strategy that combines both message-based and code-based analysis. (i) At the commit message level, we design a tailored regular expression filter to detect security-related commits based on developer-written descriptions. (ii) At the code level, we fine-tune a BERT model that performs automated semantic analysis of code changes to assess whether a commit qualifies as a security patch. The final set of security-related commits, $C_{SC}$, is obtained by integrating the results of both analyses.

\subsubsection{WFCs Extraction:}
The fourth stage is responsible for identifying the set of weakness-fixing commits (WFCs), resulting in a subset of the previously obtained security-related commit set $C_{SC}$, denoted as $C_{WFC}$, where $C_{WFC} \subseteq C_{SC}$. This process evaluates the semantic similarity between the textual message of each commit $c\in C_{SC}$ and the natural language description of each weakness $w\in W_{CWE}$, where $W_{CWE}$ denotes the set of weaknesses associated with a given CWE category. The identification process consists of two main steps. First, both commit messages and CWE weakness descriptions are transformed into high-dimensional vector representations. To ensure robustness, we apply multiple sentence embedding models, each capturing different linguistic features to encode meaning from the text. Second, for each commit $c\in C_{SC}$, we compute the cosine similarity between its vector representation and that of each weakness $w\in W$. Cosine similarity quantifies semantic alignment by measuring the cosine of the angle between two vectors, with higher values indicating stronger semantic similarity \cite{xia2015learning}. A commit $c\in C_{SC}$ is assigned to a specific CWE category $w\in W$ only if multiple models independently yield the highest similarity score for that category. This majority-voting strategy ensures consistency across embedding models. As a result, we derive the final set of WFCs, namely $C_{WFC}$.

\subsubsection{WCCs Extraction:}
The fifth stage involves identifying the set of weakness-contributing commits (WCCs) by tracing backward from each commit $c \in C_{WFC}$, denoted as $C_{WCC}$. To perform this analysis, we employ a commit-tracing algorithm that evaluates commit-level code differences to identify earlier changes that may have introduced weaknesses. 
By applying this tracing procedure, we extract the complete set of WCCs that are potentially responsible for the weaknesses present in the WFCs.

\subsubsection{Weaknesses Identification:}
While each WFC is generally linked to a distinct weakness, in practice, some weaknesses may require multiple fixes and thus be associated with several WFCs. The sixth stage aims to prevent duplicate counting of such weaknesses by introducing a deduplication and merging approach based on inclusion relationships among WCC chains. Specifically, we first examine whether WCC chains originating from different WFCs are nested and share the same earliest WFC node. If so, these chains are treated as multiple fixes for the same weakness and are merged accordingly. This yields a final set of weaknesses with duplicates eliminated, denoted as $W_{WC}$, where each weakness in $W_{WC}$ is characterized by a chain consisting of one WFC and its associated WCCs. The sets $C_{WFC}$ and $C_{WCC}$ are updated accordingly to reflect the removal of redundant entries.

\subsubsection{Manual Verification:}
The seventh stage involves manually evaluating the reliability of the identified security weaknesses. 
Given the impracticality of reviewing every weakness manually, we employ a statistical sampling method to achieve a representative evaluation. This method adheres to established inference guidelines\cite{agresti1998approximate}, using a specified confidence level, along with a predefined margin of error to determine the required sample size. Each sample contains a specific weakness identified within $W_{WC}$, forming the basis for our subsequent manual evaluation, which relies on the following information:
\begin{itemize}
    \item \textbf{Commit message:} The textual descriptions accompanying the code commit that fixes the weakness.
    \item \textbf{Contextual information:} Developers' descriptions and comments in weakness issue report.
    \item \textbf{Code diff:} The exact lines added, removed, or modified, stripped of surrounding noise.
\end{itemize}
}
To minimize subjectivity and bias in the verification process, we implement a structured, multi-step review methodology, inspired by established practices in prior studies~\cite{li2024assessing, DBLP:conf/icse/DingCS20}:
\begin{itemize}
    \item \textbf{Step 1: Dual coding.} Two authors independently analyzed the same set of weaknesses to identify inconsistencies in interpretation.
    \item \textbf{Step 2: Disagreement resolution.} In cases where the initial interpretations differ, a third author mediates a discussion to reach a consensus. 
    \item \textbf{Step 3: Iterative process:} The analysis continues until there are no new weaknesses to analyze.
\end{itemize}



\subsection{RQs Definition}

The goal of our empirical study is to investigate the characteristics and lifecycle of vulnerabilities in open-source VR software across various project types, focusing on how they are introduced, persist, and are ultimately mitigated. To achieve these objectives, we formulate nine RQs across three categories based on the VR weakness lifecycle defined in Section \ref{sec:lifecycle_definations}. 

\textbf{Overall Analysis} (RQ1$\sim$RQ3) provides a foundational understanding of the distribution and trends of different types of VR weaknesses across VR projects over time.

\begin{itemize}
    \item \textbf{RQ1: {What are the dominating types of VR weaknesses in VR software?}} We investigate the distribution of various VR weakness types across different project types, where the VR weaknesses are classified according to the CWE, as previously discussed. Addressing this question will reveal which VR weaknesses are more common in specific project types, thereby assisting researchers and developers in prioritizing their efforts.

    \item \textbf{RQ2: How are VR weaknesses distributed among various VR software types?} We examine the distribution of VR weaknesses by analyzing both the number and density of weaknesses across various VR project types, using total weakness counts and file sizes as key metrics. This analysis reveals which types of projects are more susceptible to security issues and provides actionable insights for developers to implement targeted mitigation strategies.

    \item \textbf{RQ3: How do VR weaknesses in VR software evolve over time?} 
    We track the annual occurrence of various typical VR weakness types to identify patterns in weakness introduction, thereby highlighting whether certain VR weaknesses are increasing or decreasing in prevalence. This temporal analysis offers valuable insights into the changing landscape of VR weaknesses, enabling developers to adapt to the evolving weakness profile.
\end{itemize}

\textbf{VR weakness Contribution Analysis} (RQ4$\sim$RQ6) explores the conditions and factors involved in the introduction of vulnerabilities.

\begin{itemize}
    \item \textbf{RQ4: When are VR weaknesses introduced into VR software, and how long does the introduction process typically take?} We examine whether VR weaknesses are introduced during the initial file creation phase or emerge through subsequent maintenance activities. For weaknesses introduced during creation, we further analyze which file types are more susceptible to introducing them. In addition, we investigate the insertion window of the leading CWE types to understand which categories tend to have more complex or prolonged introduction processes, offering deeper insights into their underlying development dynamics.

    \item \textbf{RQ5: How are VR weaknesses introduced into VR software?} We examine several factors that may influence the introduction of VR weaknesses, including commit goals, commit frequency, insertion windows, and changes to both code and files. 
    We further investigate whether VR weaknesses are introduced through third-party libraries. Understanding this aspect helps pinpoint the sources of weaknesses and enables the development of more targeted and effective prevention strategies in VR software.

    \item \textbf{RQ6: How does developer status affect VR weaknesses introduction?} We analyze the impact of developer experience and workload on the number of WCCs within the VR project, with the goal of understanding whether experienced developers are less likely to introduce VR weaknesses and how increased workload or pressure impacts the introduction of weaknesses. This could result in recommendations for more rational software development planning and task distribution, reducing the likelihood of introducing VR weaknesses.

\end{itemize}

\textbf{VR weakness Survival Analysis} (RQ7$\sim$RQ9) investigates the longevity of VR weaknesses in the repository and the methods utilized to resolve them. 

\begin{itemize}
    \item \textbf{RQ7: How do the lifetimes of different VR weaknesses vary in diverse contexts?} We measure the impact of weakness type and project type on the survival duration of each weakness, and analyze the detection delay for each. Answering this question helps pinpoint which weakness types persist longer, enabling developers to prioritize the resolution of specific VR weaknesses.
    
    \item \textbf{RQ8: How are VR weaknesses removed from the source code?} We investigate the typical approaches used to address VR weaknesses, analyzing the complexity and efficiency of the fixes. Answering this question allows us to optimize the weakness resolution process and inform best practices for effective and efficient fixes. 

    {
    \item \textbf{RQ9: How effective are current code analysis tools in detecting VR weaknesses in VR software?} We examine the challenges encountered by dynamic analysis techniques in detecting VR weaknesses in VR software. In parallel, we apply the widely adopted static analysis tool CodeQL to assess weaknesses in representative projects. This investigation reveals the limitations of current automated detection tools in the VR domain and identifies avenues for enhancing their effectiveness.
    }
\end{itemize}

\section{Our Dataset}
\label{sec:dataset_collection}
{\color{black}
We obtain our dataset required for this empirical study by implementing the dataset construction method of the proposed framework, as outlined in Section \ref{sec:dataset_construction}.

\subsection{Projects Selection}
To identify the VR-related GitHub repositories suitable for our analysis, we first reference the benchmark dataset from the study by Rzig et al. \cite{Rzig2023Virtual} published in 2023, containing git clone links for 315 manually-verified VR open-source projects. After excluding 5 projects that are no longer accessible and 4 with fewer than 100 commits due to recent updates or re-uploads, we retain 306 projects. Additionally, we expand our dataset to include the latest projects created on GitHub between January 1, 2023, and December 31, 2024, by searching for new VR projects using keywords like ``VR'', ``Virtual Reality'', and ``VR Unity'', while ensuring that each project has at least 100 historical commits. This process leads to the identification of 28 new projects that meet the inclusion criteria but are not part of the original dataset. Through the combined use of both methods, we derive the set of qualified projects, namely $P$, consisting of 334 projects in total.

Among the total 334 projects, 125 are development tools, including 25 software development kits, 3 graphics and rendering engines, 28 development frameworks, 4 drivers, 53 plugin tools, and 12 tutorial projects, while the remaining 209 are VR applications, comprising 82 games, 94 tools, and 33 modules. On average, these projects are developed and maintained by 10 authors. 
The longest-running project \verb|vrs|\footnote{\url{https://github.com/vradarserver/vrs}} contains 1,387 commits over 3,332 days, and the shortest project \verb|vroom|\footnote{\url{https://github.com/vihanchaudhry/vroom}} has 179 commits over just 1 day. Most projects are written in C\#, comprising 295 C\#-based projects (88.3\%), with a few in C++, JavaScript, Python, etc.

\subsection{Commits Extraction}
We locally clone each of the 334 projects and extract their commit histories using git commands and GitHub API~\cite{GitHub-github_api-11-12}. This process results in the complete set of commits across all repositories, namely $C$, comprising a total of 183,259 commits, with an average of 548 commits per project and a median of 255 commits. The project with the highest commit count is \verb|MixedRealityToolkit-Unity|\footnote{\url{https://github.com/microsoft/MixedRealityToolkit-Unity}}, a Microsoft project designed to provide a comprehensive toolkit for developing VR applications in Unity. It is active for 2,148 days from its first commit on February 23, 2018, accumulating 17,118 commits.

\subsection{Security-related Commits Extraction}
To identify the set of security-related commits $C_{SC}$ from the complete set $C$, we perform both message-based and code-based analysis. At the commit message level, we apply a filter based on weakness-related keywords. The filter explicitly excludes common non-security-related terms associated with routine maintenance or functional updates (e.g., ``merge'',  ``test'', ``configuration''), while actively capturing terms indicative of security issues, including remedial actions (e.g., “fix”, “patch”, “resolve”), references to weaknesses (e.g., “vulnerability”, ``bug'', “weakness”, “exposure”, “threat”), and specific weakness types (e.g., “SQL injection”, “XSS”).

At the commit code level, we utilize the LLMDA framework ~\cite{Tang2024Just} and fine-tune a Transformer-based BERT model using PyTorch 1.13.0 on an NVIDIA RTX 3060 GPU (16GB, CUDA 11.6). The model receives inputs from \verb|git-diff|, including code patches, file names, and change types. It is trained for 20 epochs, following the setup from the original study \cite{Tang2024Just}, and is then applied to classify commits as either security-related or non-security-related.

Using both methods, we identify 18,608 security-related commits from the total set of 183,259 commits ($C$), comprising the set $C_{SC}$. This number is reasonable, considering that many commits in VR software development are dedicated to non-security tasks like documentation, configuration, and asset management. The average number of security-related commits per project is 55, with a median of 21, a maximum of 2,572, and a minimum of 1.

\subsection{WFCs Extraction}
Recall that weakness-fixing commits (WFCs) are identified by evaluating the semantic similarity between the textual descriptions of each commit message and specific security weaknesses in the CWE. For this purpose, we choose CWE-699 to classify VR-related weaknesses, as it serves as a parent category encompassing 40 typical software development weaknesses, thus $|W|=40$.
The descriptions of both commit messages and CWE weaknesses are transformed into high-dimensional vector representations using five distinct sentence embedding models.

Table.~\ref{tab:pretrained_models} shows the five sentence-to-vector models utilized in this study, including a traditional method and four advanced pre-trained models \cite{HuggingFace}. Model 1 produces weighted sentence vectors by computing term frequency-inverse document frequency (TF-IDF) \cite{salton1983introduction} metrics, effectively capturing salient keywords and their relative importance within texts. 
{A detailed description of Model 1 is provided in Appendix A in the supplemental file.} 
In contrast to Model 1, Models 2-5 leverage pre-trained architectures trained on extensive text corpora, generating contextualized embeddings that capture syntactic relationships and semantic nuances, enabling more sophisticated text interpretation. 

After constructing vector spaces for both commit messages and weaknesses within CWE-699, we employ cosine similarity to evaluate their semantic similarity and identify potential WFCs. For each commit, we compute its semantic similarity to the 40 weakness descriptions using the five models.
A commit is assigned to a specific CWE category only if more than four models assign it the highest similarity score for that category. Otherwise, discrepancies are resolved through
manual analysis. Commits that receive a classification are treated as WFCs.

As a result of this process, we identify 2,703 WFCs from 18,608 security-related commits ($C_{SC}$), which form the set $C_{WFC}$, with an average of 8 WFCs per project, a maximum of 357, a minimum of 0, and a median of 3. Notably, 49 projects yield no CWE-classified weaknesses, due to disagreement among the outputs of the five models.

\begin{table}[!htbp]
    \centering
    \caption{Models Used in Sentence Vector Construction}
    \label{tab:pretrained_models}
    \resizebox{\linewidth}{!}{
        \begin{threeparttable}
        \begin{tabular}{c|c}
            \hline
            Model No. & Model Name \\
            \hline
            Model 1 & TF-IDF + Word2Vec\tnote{1} \\
            Model 2 & all-MiniLM-L6-v2\tnote{2} \\
            Model 3 & basel/ATTACK-BERT\tnote{3} \\
            Model 4 & sentence-transformers/all-mpnet-base-v2\tnote{4} \\
            Model 5 & sentence-transformers/paraphrase-multilingual-mpnet-base-v2\tnote{5} \\
            \hline
        \end{tabular}
        \begin{tablenotes}
            \item [1] See in Appendix A in the supplemental file.
            \item[2] \url{https://huggingface.co/sentence-transformers/all-MiniLM-L6-v2}
            \item[3] \url{https://huggingface.co/basel/ATTACK-BERT}
            \item[4] \url{https://huggingface.co/sentence-transformers/all-mpnet-base-v2}
            \item[5] \url{https://huggingface.co/sentence-transformers/paraphrase-multilingual-mpnet-base-v2}
        \end{tablenotes}
        \end{threeparttable}
    }
    \vspace{-0.5cm}
\end{table}

\subsection{WCCs Extraction}
We determine the weakness-contributing commits (WCCs) corresponding to each weakness-fixing commit (WFC) in $C_{WFC}$ by employing a commit-tracing algorithm inspired by the SZZ approach \cite{sliwerski2005changes}, which explores the version histories to uncover earlier changes that might have led to the weaknesses later addressed. In our implementation, we extend the open-source SZZ algorithm by \cite{Iannone2023Secret}, incorporating a multiprocessing enhancement to improve scalability and efficiency for large-scale commit tracing. From the 2,703 WFCs analyzed, 1,708 are successfully traced, resulting in 7,658 WCCs that make up the set $C_{WCC}$.

\subsection{Weaknesses Identification}
Through the WCCs extraction, each traceable WFC is linked to a corresponding chain of WCCs. To avoid counting multiple WFCs that address the same issue as separate weaknesses, we apply a deduplication process that identifies and merges WFCs linked to the same underlying weakness by analyzing inclusion relationships among their WCC chains. As illustrated in Figure~\ref{fig:deduplication}, the WCC chains of WFC-\verb|79373| and WFC-\verb|76142| are contained within the WCC chain of WFC-\verb|cce8d|. The earliest WCC of Weakness I (i.e., WCC-\verb|c4ale|) is identical to that of Weakness II, indicating that WFC-\verb|79373| and WFC-\verb|cce8d| address the same weakness. Therefore, we incorporate Weakness I as a historical fix record into Weakness II and remove Weakness I from the dataset. In contrast, Weakness III has a different earliest WCC (i.e., WCC-\verb|f8eld|), suggesting that WFC-\verb|cce8d| and WFC-\verb|76142| resolve distinct weaknesses. As a result, we retain Weakness III. 

The process ultimately identifies 27 weaknesses that have been addressed multiple times, distributed across 20 distinct projects. Duplicates are removed and merged into existing weaknesses. In the end, we obtain a dataset containing 1,681 unique weaknesses, which forms the set $W_{WC}$, along with 1,708 WFCs, and 7,595 WCCs.

\begin{figure}[!htbp]
    \centering
    \includegraphics[width=0.8\linewidth]{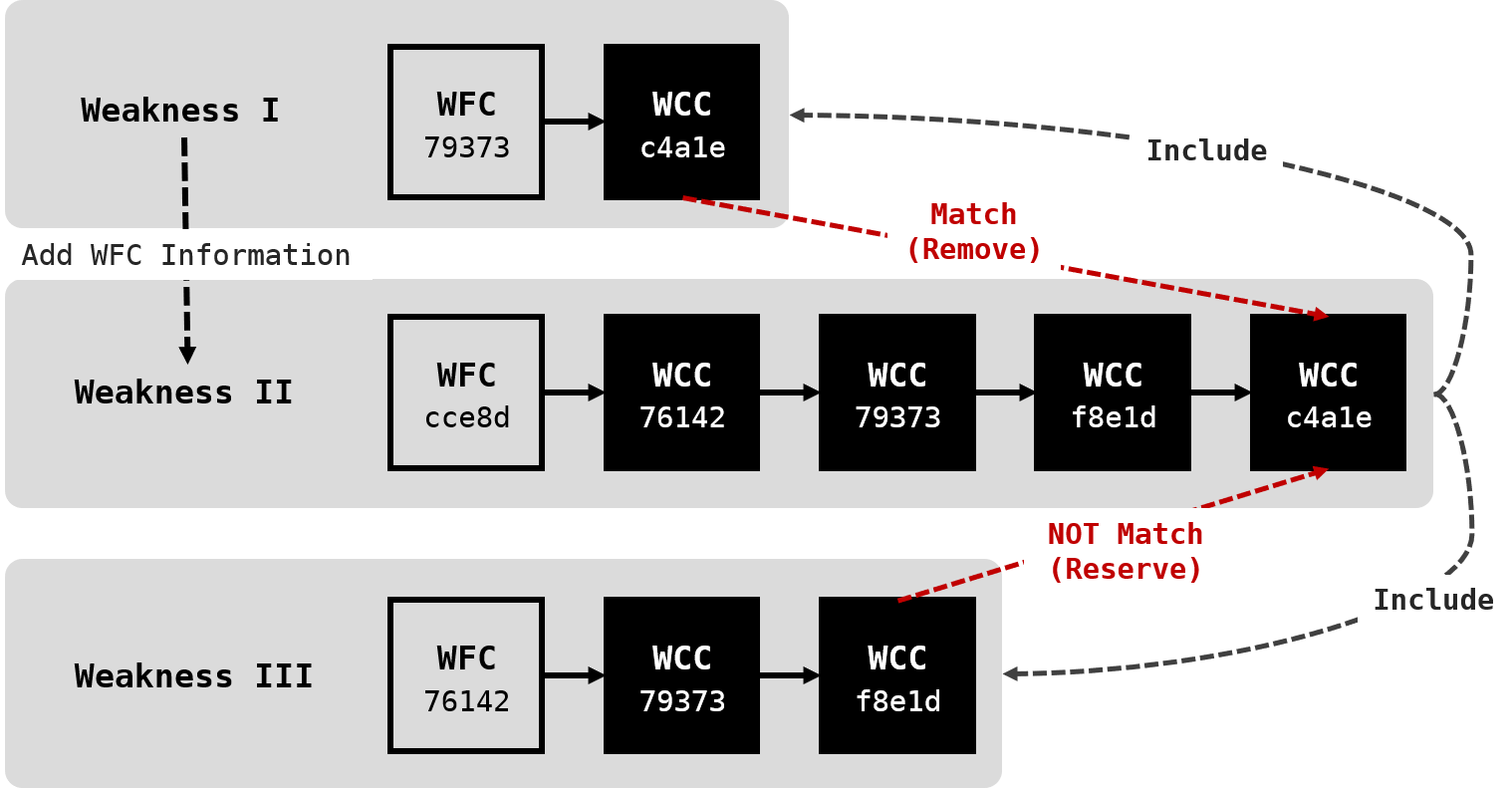}
    \caption{Weakness deduplication process.}
    \label{fig:deduplication}
    \vspace{-0.5cm}
\end{figure}

\subsection{Manual Verification}
To verify the reliability of the obtained 1,681 VR weaknesses, we conduct a rigorous manual verification process. Using a standard formula for unknown populations~\cite{daniel2018biostatistics}, we determine that a sample size of 318 weaknesses would provide a 95\% confidence level with a ±5\% margin of error. This approach balances analytical precision with practical feasibility, ensuring our findings are both statistically valid and operationally manageable. We randomly sample 318 weaknesses from the 1,681 weaknesses collected in the previous step. The verification involves manually examining various elements of each weakness
to validate both its existence and the correctness of its CWE classification. By employing a three-step review process involving dual coding, disagreement resolution and iteration, the analysis achieves a Cohen's Kappa score of 88\% between the two authors, indicating an almost perfect agreement~\cite{mchugh2012interrater}.

Among the 318 randomly sampled weaknesses, 309 weaknesses (97.2\%) are confirmed as valid with correct CWE labels, yielding an estimated overall accuracy of 97\% for the full dataset, with a 95\% confidence interval of $95.3\%\sim98.7\%$.

\section{Empirical Analysis}
\label{sec:evaluation}

\newcounter{insightcounter}

\newcommand{\insightbox}[1]{%
    \refstepcounter{insightcounter}
    \vspace{0.5em} 
    \textbf{\textsc{Insight \theinsightcounter.}}\hspace{0.5em}{\em #1}\par
    \vspace{0.5em} 
}

\subsection{Overall Analysis}

\subsubsection{RQ1: What are the dominating types of VR weaknesses in VR software?}
\label{sec:rq1_answer}
\ 
\newline
\indent 
{\color{black}
Recall that the 1,681 weaknesses we identified fall into 40 distinct CWE types defined under the CWE-699 hierarchy. Table \ref{tab:top_20_cwe} lists the top 20 CWE types and their corresponding weakness counts, which together make up more than 93\% of all identified VR weaknesses. We observe that, unlike traditional software, where issues like \textit{Data Neutralization Issues} (CWE-137) and \textit{Memory Buffer Errors} (CWE-1218) are more common  \cite{shi2022does, Iannone2023Secret}, VR software exhibits a higher proportion of user-related weaknesses, such as \textit{User Interface Security Issues} (CWE-355). This increased prevalence of UI-related weaknesses in VR software likely stems from its reliance on continuous, real-time user interactions within a first-person perspective, coupled with the added complexity of integrating multiple hardware devices and configuration files.

\begin{table*}[t]
\caption{Statistics of Top 20 CWE Types}
    \centering
    \resizebox{\textwidth}{!}{%
    \begin{tabular}{c c c c || c c c c}
    \hline
         \textbf{No.} & \textbf{CWE ID} & \textbf{Name} & \textbf{Number} &  \textbf{No.} & \textbf{CWE ID} & \textbf{Name} & \textbf{Number} \\
    \hline 
         1 & 355 & User Interface Security Issues & 657 & 11 & 411 & Resource Locking Problems & 36 \\
         2 & 1219 & File Handling Issues & 129 & 12 & 429 & Handler Errors & 30 \\
         3 & 465 & Pointer Issues & 124 & 13 & 1217 & User Session Errors & 30 \\
         4 & 569 & Expression Issues & 94 & 14 & 19 & Data Processing Errors & 29 \\
         5 & 133 & String Errors & 88 & 15 & 317 & State Issues & 23 \\
         6 & 387 & Signal Errors & 70 & 16 & 1215 & Data Validation Issues & 20 \\
         7 & 189 & Numeric Errors & 65 & 17 & 1213 & Random Number Issues & 18 \\
         8 & 1218 & Memory Buffer Errors & 41 & 18 & 137 & Data Neutralization Issues & 15 \\
         9 & 1210 & Audit / Logging Errors & 37 & 19 & 1225 & Documentation Issues & 15 \\
         10 & 136 & Type Errors & 36 & 20 & 275 & Permission Issues & 14 \\
    \hline
    \end{tabular}
    }
    \label{tab:top_20_cwe}
    \vspace{-0.1cm}
\end{table*}

One instance of \textit{User Interface Security Issues} (CWE-355) is observed in the development tool project \verb|wrapVR|\footnote{\url{https://github.com/mynameisjohn/wrapVR}}, particularly in the commit identified by the short hash \verb|25a2b4d|. As illustrated in Figure~\ref{fig:cwe_example}, this update to the \verb|Grabbable.cs| file addresses a UI component access issue. The previous implementation attempted to access a UI component without validating its existence, posing a risk of null reference exceptions or unintended behavior if the component was missing. The revised code introduces a validation mechanism that checks the component's presence before proceeding with interaction callbacks, thereby improving stability and minimizing unintended behavior in user interactions.

\begin{figure}[!htbp]
    \centering
    \includegraphics[width=0.8\linewidth]{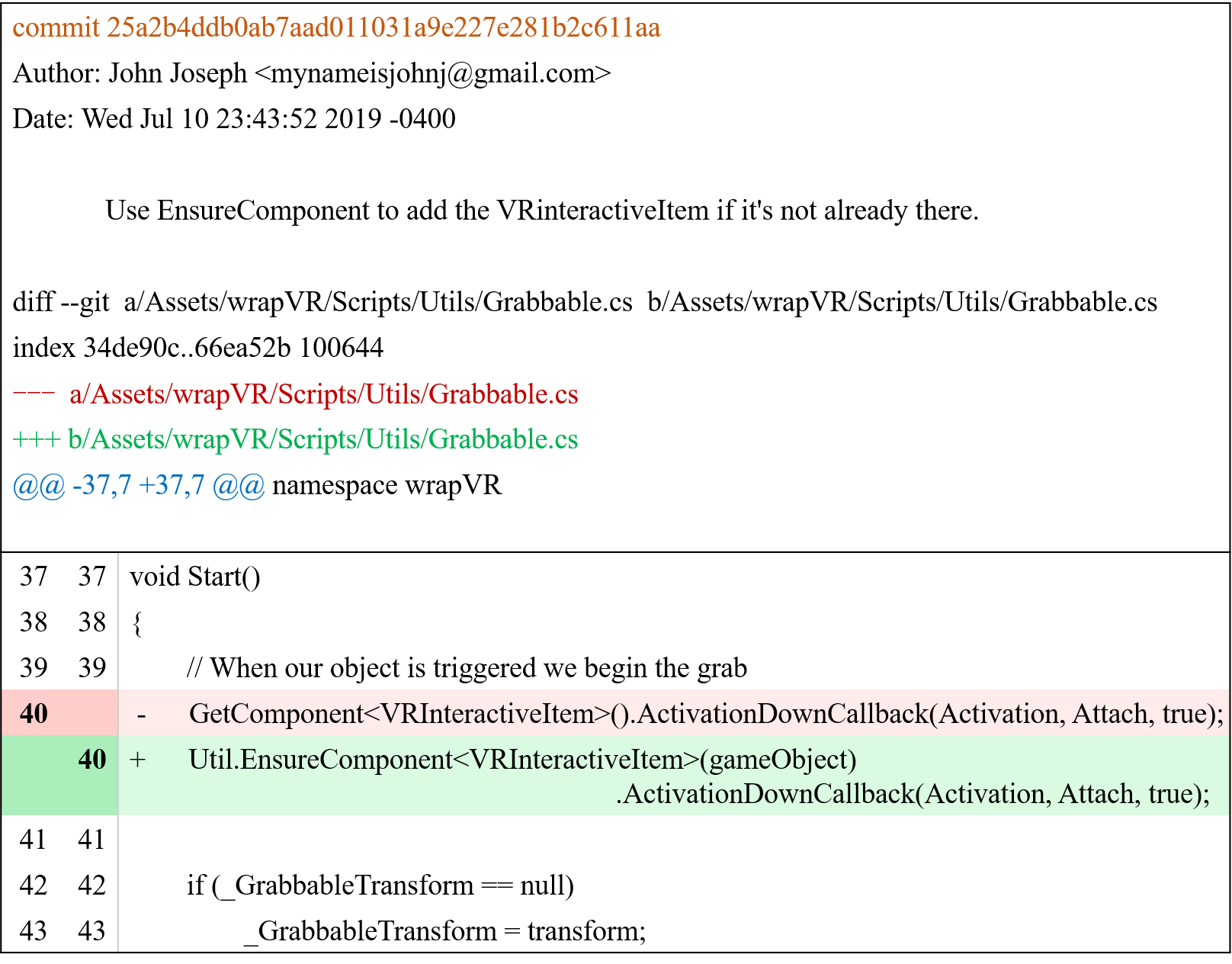}
    \caption{An example of CWE-355 in a VR game project.}
    \label{fig:cwe_example}
    \vspace{-0.3cm}
\end{figure}

Additionally, VR weaknesses related to resource handling, such as \textit{File Handling Issues} (CWE-1219) and \textit{Pointer Error} (CWE-465), are also frequently observed in VR software. This is likely due to the substantial resource demands of VR systems, which require real-time processing of large volumes of data, including 3D graphics, user input, and sensory feedback. 
The necessity of synchronizing numerous tasks in real time increases the risk of errors related to memory management and data processing. Moreover, the rapid iteration of VR engines and frequent updates to tools like 3D modeling software can introduce compatibility issues, further elevating the risk of file handling and pointer errors during asset integration.

\insightbox{VR weaknesses are dominated by user interface weaknesses
like CWE-355 (39.0\%), followed by resource-related weaknesses like CWE-1219 (7.6\%) and CWE-465 (7.4\%).}

\subsubsection{RQ2: How are VR weaknesses distributed among various VR software types?}
\label{sec:rq2_answer}
\ 
\newline
\indent 
By examining the project descriptions and implementations, we divide our dataset into two primary classes—applications and development tools—comprising nine categories. To find the distribution of VR weaknesses across various categories of VR projects, we measure the total project size and total weaknesses within each category, and express weakness density as the ratio of these two values, as summarized in Table~\ref{tab:memory_per_class}.

\begin{table}[!htbp]
\caption{Statistics on Security Weaknesses by Category}
\resizebox{\columnwidth}{!}{%
    \begin{tabular}{c|c|c|c|c}
    \hline
    \textbf{Class} & \textbf{Category} & \textbf{Size (GB)} & \textbf{\makecell{Weakness \\ count}} & \textbf{\makecell{Weakness \\ density}} \\ \hline
    \multirow{3}{*}{Application} & Game & 127.82 & 96 & 0.75 \\ \cline{2-5} 
     & Module & 144.34 & 287 & 1.99 \\ \cline{2-5} 
     & {Utility} & 14.22 & 163 & 11.46 \\ \hline
    \multirow{6}{*}{\makecell{Development \\ Tool}} & {SDK} & 10.33 & 319 & 30.88 \\ \cline{2-5} 
     & {Plugin} & 13.55 & 427 & 31.51 \\ \cline{2-5} 
     & {Tutorial} & 21.22 & 18 & 0.85 \\ \cline{2-5} 
     & {Development Framework} & 11.49 & 336 & 29.24 \\ \cline{2-5} 
     & {Driver} & 3.11 & 3 & 0.96 \\ \cline{2-5} 
     & {Graphics Engine} & 0.42 & 32 & 76.19 \\ \hline
    \end{tabular}
    }
    \label{tab:memory_per_class}
    \vspace{-0.2cm}
\end{table}

We find that the total file size of development tools (60.12GB) is much smaller than that of application software (286.38GB), yet development tools generate far  more weaknesses. In terms of absolute weakness count, development tools contain 1,135 weaknesses, while application software contains 536, which is 2.12 times higher. In terms of relative weakness density, development tools have a density of 169.63, compared to 14.2 for applications, which is 11.95 times higher. This suggests that development tools carry much higher security risks than applications.

When considering each individual category, plugin exhibits the highest number of weaknesses (427), followed by development framework (336) and SDK (319). Meanwhile, graphics engine shows the highest weakness density (76.19), followed by plugin (31.51), SDK (30.88), and development framework (29.24). This necessitates enhanced security measures and proactive monitoring for these categories of VR software.

\insightbox{VR development tools are smaller than VR applications but incur higher security risks in terms of both the number and density of weaknesses, with plugin, SDK, development framework, and graphics engine being particularly vulnerable.}

\begin{table*}[t]
\caption{Top 10 CWE Types of Different Project Categories}
\resizebox{\textwidth}{!}{%
    \begin{tabular}{c|c||cccccccccc|c}
        \hline
        \textbf{Class} & \textbf{Category} & \textbf{CWE-355} & \textbf{CWE-1219} & \textbf{CWE-465} & \textbf{CWE-569} & \textbf{CWE-133} & \textbf{CWE-387} & \textbf{CWE-189} & \textbf{CWE-1218} & \textbf{CWE-1210} & \textbf{CWE-136} & \textbf{Sum} \\
        
        \hline
        \multirow{6}{*}{\textbf{Application}} & \multirow{2}{*}{\textbf{Game}} & \textbf{47} & 4 & 1 & 3 & 2 & \textbf{9} & 3 & 2 & 2 & 0 & 73 \\
        &  & \textbf{64.38\%} & 5.48\% & 1.37\% & 4.11\% & 2.74\% & \textbf{12.33\%} & 4.11\% & 2.74\% & 2.74\% & 0.00\% & 76.04\% \\
        \cline{2-13}
        & \multirow{2}{*}{\textbf{Module}} & \textbf{90} & 9 & 1 & 5 & 6 & \textbf{10} & 5 & 1 & 6 & 0 & 133 \\
        &  & \textbf{67.67\%} & 6.77\% & 0.75\% & 3.76\% & 4.51\% & \textbf{7.52\%} & 3.76\% & 0.75\% & 4.51\% & 0.00\% & 81.60\% \\
        \cline{2-13}
        & \multirow{2}{*}{\textbf{\makecell{Utility}}} & \textbf{91} & \textbf{48} & 8 & 10 & 13 & 11 & 11 & 11 & 4 & 4 & 211 \\
        &  & \textbf{43.13\%} & \textbf{22.75\%} & 3.79\% & 4.74\% & 6.16\% & 5.21\% & 5.21\% & 5.21\% & 1.90\% & 1.90\% & 73.52\% \\
         
        \hline
         
        \multirow{12}{*}{\textbf{\makecell{Development \\ Tool}}} & \multirow{2}{*}{\textbf{SDK}} & \textbf{83} & 19 & 15 & \textbf{30} & 21 & 19 & 17 & 12 & 15 & 5 & 236 \\
         &  & \textbf{35.17\%} & 8.05\% & 6.36\% & \textbf{12.71\%} & 8.90\% & 8.05\% & 7.20\% & 5.08\% & 6.36\% & 2.12\% & 73.98\% \\
         \cline{2-13}
         & \multirow{2}{*}{\textbf{Plugin}} & \textbf{211} & 22 & \textbf{44} & 20 & 30 & 11 & 15 & 5 & 6 & 6 & 370 \\
         &  & \textbf{57.03\%} & 5.95\% & \textbf{11.89\%} & 5.41\% & 8.11\% & 2.97\% & 4.05\% & 1.35\% & 1.62\% & 1.62\% & 86.65\% \\
         \cline{2-13}
         & \multirow{2}{*}{\textbf{Tutorial}} & \textbf{16} & 1 & 0 & 0 & 0 & 0 & 0 & 1 & 0 & 0 & 18 \\
         &  & \textbf{88.89\%} & 5.56\% & 0.00\% & 0.00\% & 0.00\% & 0.00\% & 0.00\% & 5.56\% & 0.00\% & 0.00\% & 100.00\% \\
         \cline{2-13}
         & \multirow{2}{*}{\textbf{\makecell{Development \\ Framework}}} & \textbf{105} & 25 & \textbf{55} & 23 & 15 & 10 & 11 & 9 & 4 & 13 & 270 \\
         &  & \textbf{38.89\%} & 9.26\% & \textbf{20.37\%} & 8.52\% & 5.56\% & 3.70\% & 4.07\% & 3.33\% & 1.48\% & 4.81\% & 80.36\% \\
         \cline{2-13}
         & \multirow{2}{*}{\textbf{Driver}} & \textbf{1} & 0 & 0 & 0 & \textbf{1} & 0 & \textbf{1} & 0 & 0 & 0 & 3 \\
         &  & \textbf{33.33\%} & 0.00\% & 0.00\% & 0.00\% & \textbf{33.33\%} & 0.00\% & \textbf{33.33\%} & 0.00\% & 0.00\% & 0.00\% & 100.00\% \\
         \cline{2-13}
         & \multirow{2}{*}{\textbf{\makecell{Graphics \\ Engine}}} & \textbf{13} & 1 & 0 & 3 & 0 & 0 & 2 & 0 & 0 & \textbf{8} & 27 \\
         &  & \textbf{48.15\%} & 3.70\% & 0.00\% & 11.11\% & 0.00\% & 0.00\% & 7.41\% & 0.00\% & 0.00\% & \textbf{29.63\%} & 84.38\% \\
         
         \hline
    \end{tabular}%
    }
    \label{tab:project_type_cwe}
\end{table*}

We further analyze the distribution of CWE types across various project categories within application and development tool classes, with the results summarized in Table \ref{tab:project_type_cwe}. This table presents a detailed breakdown of the top 10 CWE weakness types derived from Table \ref{tab:top_20_cwe} for each category. Each cell displays the raw number of weaknesses alongside their percentage relative to the total, with bold values highlighting the CWE types with the highest and second-highest proportions. The final column, labeled “Sum”, aggregates the total counts and percentages of these CWE types relative to the overall weaknesses identified per category.

We observe that both applications and development tools are predominantly affected by \textit{User Interface Security Issues} (CWE-355), suggesting that intensive user interaction is a fundamental source of security risk across different VR contexts. In terms of secondary issues, notable differences exist not only between applications and development tools, but also among various categories within each class.

Specifically, among applications, game and module types often encounter \textit{Single Errors} (CWE-387), likely due to their complex and dynamic state transitions, where rapid processing of sensor data increases the likelihood of mishandling exceptional conditions. Conversely, utility tools are particularly prone to \textit{File Handling Issues} (CWE-1219), likely resulting from frequent file operations like media file loading, which elevate the risk of input errors and resource leaks. For development tools, aside from the dominance of \textit{User Interface Security Issues} (CWE-355), other relatively frequent weaknesses vary significantly across categories, reflecting the architectural heterogeneity of such tools. Among them, \textit{Pointer Errors} (CWE-465) and \textit{Expression Issues} (CWE-569) occur more frequently than other types.

\insightbox{
While both VR applications and VR development tools predominantly face user interface-related weaknesses, the former are more prone to file and signal errors, and the latter exhibit greater susceptibility to memory and logic faults, highlighting the need for differentiated, context-aware mitigation measures.
}

\subsubsection{RQ3: How do VR weaknesses in VR software evolve over time?}
\label{sec:rq3_answer} 
\ 
\newline
\indent 
To answer this RQ, we analyze the longitudinal trend of VR weaknesses from 2014 to 2024. As shown in Table \ref{tab:vulnerability_trend}, the number of VR weaknesses and the weaknesses-to-commit ratio fluctuate during this period, 
peaking between 2016 and 2018, followed by a decline and a gradual resurgence in recent years. This trajectory aligns with the cycles of VR technology acceleration and market expansion. The shipment of 6.3 million headsets in 2016 alone \cite{SuperData:-2025-02-28, VR-2025-02-28} exemplifies this boom, during which functionality and speed-to-market were prioritized over security \cite{Martin2017Virtual, hall2017augmented, Abbasi2017Technology}. Since 2019, the stabilization of the VR market and the adoption of preliminary security measures have contributed to a slight reduction in software weaknesses. In the past two years, the growing integration of VR with other domains and increased system complexity have led to a renewed rise in security issues.

\begin{table*}
    \caption{Annual Security Weaknesses Trends of All CWE Types}
    \centering
    \resizebox{\textwidth}{!}{%
    \begin{tabular}{c|ccccccccccc}
        \hline
        \textbf{Year} & \textbf{2014} & \textbf{2015} & \textbf{2016} & \textbf{2017} & \textbf{2018} & \textbf{2019} & \textbf{2020} & \textbf{2021} & \textbf{2022} & \textbf{2023} & \textbf{2024} \\
        \hline
        \textbf{Commits} & 1161 & 4772 & 17635 & 19777 & 19367 & 24133 & 22902 & 19705 & 18091 & 19448 & 16500 \\
        \textbf{Vulnerabilities} & 8 & 34 & 162 & 183 & 197 & 212 & 191 & 166 & 157 & 179 & 192 \\
        \textbf{Ratio} & 6.89‰ & 7.12‰ & 9.18‰ & 9.25‰ & 10.17‰ & 8.78‰ & 8.33‰ & 8.42‰ & 8.67‰ & 9.20‰ & 11.63‰ \\
        \hline
    \end{tabular}
    }
    \label{tab:vulnerability_trend}
    \vspace{-0.3cm}
\end{table*}

\insightbox{
Driven by the surge in VR market between 2016 and 2018, VR weaknesses have increased substantially,highlighting the importance of developing domain-specific, forward-looking security solutions.}

\begin{figure}[!htbp]
    \centering
    \includegraphics[width=\linewidth]{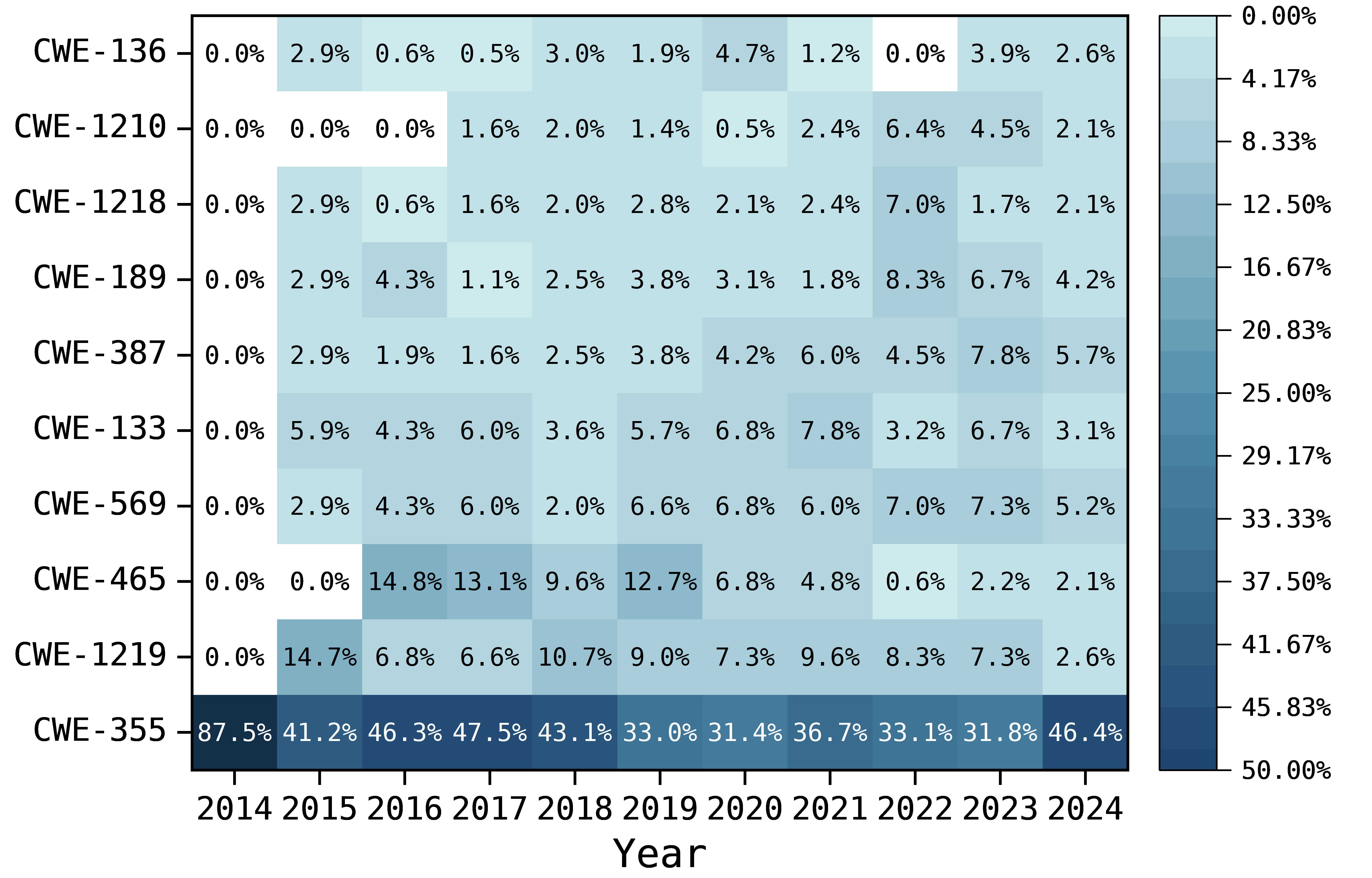}
    \caption{Annual trends of top 10 CWE types from 2014 to 2024.}
    \label{fig:characteristic_trends}
    \vspace{-0.3cm}
\end{figure}

We further explore the evolution of each CWE type over time. Figure \ref{fig:characteristic_trends} illustrates the annual proportion of VR weaknesses associated with the top 10 CWE types, relative to the total VR weaknesses reported per year. Disregarding a few outliers, we observe an overall declining trend in some VR weaknesses including \textit{User Interface Security Issues} (CWE-355), \textit{File Handling Issues} (CWE-1219), and \textit{Pointer Errors} (CWE-465), which have historically accounted for the largest proportion of weaknesses. This reflects an increased awareness of these high-risk issues and the potential effectiveness of applied security interventions. 
In contrast, some less prevalent weaknesses such as \textit{Expression Issues} (CWE-569), \textit{Data Validation Issues} (CWE-387), \textit{Numeric Errors} (CWE-189), and \textit{Audit/Logging Errors} (CWE-1210) have shown a consistent or gradual increase, primarily related to data handling. This highlights the escalating difficulties in processing sensitive data within increasingly complex dynamic multi-user systems. The remaining weaknesses like \textit{Type Errors} (CWE-136) and \textit{String Errors} (CWE-133) exhibit a relatively stable trajectory, interspersed with occasional fluctuations.

\insightbox{
In recent years, several prevalent types of VR weaknesses (e.g., CWE-355, CWE-1219, CWE-465) have been alleviated to some extent, while some previously inconspicuous weaknesses (e.g., CWE-569, CWE-387, CWE-189, CWE-1210) have shown an upward trend.}

\vspace{-0.3cm}
\subsection{Security Weaknesses Contribution Analysis}

\subsubsection{RQ4: When are VR weaknesses introduced into VR software, and how long does the introduction process typically take?}
\label{sec:rq4_answer}
\ 
\newline
\indent 
To answer this RQ, we consider the introduction of VR weaknesses at both the weakness level and the file level. At the weakness level, we examine each VR weaknesses and its associated files, whereas at the file level, we investigate each file and its related VR weaknesses, to identify when VR weaknesses are introduced.

For each of the 1,681 VR weaknesses, we examine the creation time of the file containing the weakness and the time of the first WCC associated with the weakness. 
Note that a total of 95 VR weaknesses are excluded from the analysis because their file creation timestamps could not be retrieved from the repository.
We observe that 998 (62.93\%) VR weaknesses are introduced during file creation and 566 (35.69\%) VR weaknesses appear during maintenance, while the remaining 22 (1.38\%) VR weaknesses are related to both file creation and subsequent maintenance. In addition, 2,321 files are involved in the introduction of the weakness. Among these files, 1,588 (70.23\%) files introduce VR weaknesses when they are created, and 673 (29.77\%) files introduce VR weaknesses during the maintenance process. 
The high incidence of security weaknesses during file creation suggests that VR software developers prefer overwriting entire files to making incremental modifications.

To examine the connection between file types and the VR weaknesses they introduce, we further analyze the 1,588 files where VR weaknesses are introduced at the time of file creation. The results show that \verb|C#| files account for 86.65\% of all VR weaknesses, confirming the predominant dependence on the Unity engine in the VR software development ecosystem. Smaller but notable proportions are attributed to \verb|.py| files (1.13\%) and \verb|.dwlt| files (0.69\%). These file types are commonly involved in runtime behavior scripting and scenario configuration, rendering them particularly susceptible to VR weaknesses.

\insightbox{VR weaknesses are often introduced at the VR software birth time or when the software is introduced (62.93\%) than subsequent maintenance (35.69\%), emphasizing the importance of early-stage security practices.}

\begin{figure*}[!htbp]
    \centering
    \includegraphics[width=0.9\linewidth]{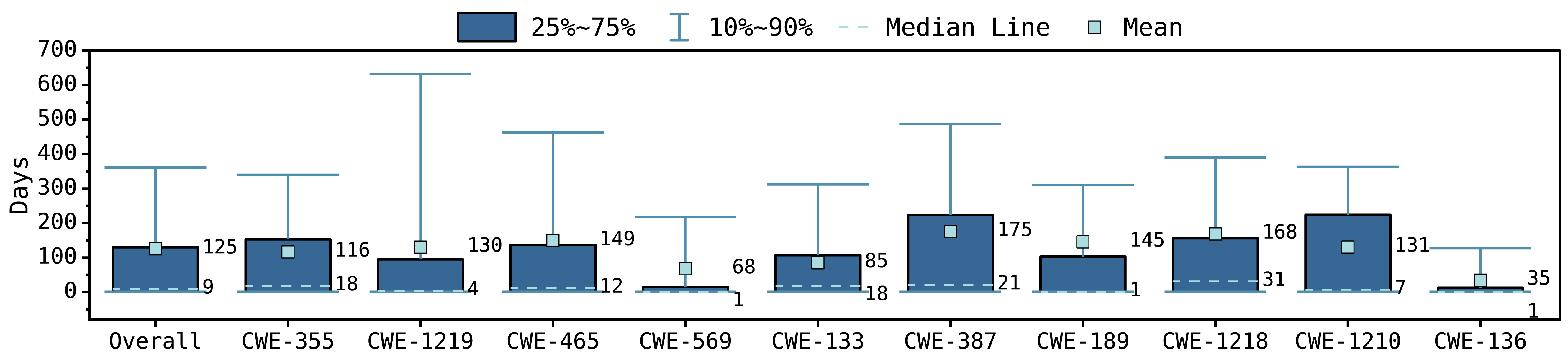}
    \caption{The insertion windows with respect to different CWE types.}
    \label{fig:insertion_window}
    \vspace{-0.5cm}
\end{figure*}

We continue to investigate the insertion windows (\(t_{0,1}\)) for various CWE types, recalling that the insertion window metric denotes the temporal range within which a VR weakness is introduced in the repository. 
The first box plot (overall) in Figure \ref{fig:insertion_window} depicts the overall time duration for all CWE types, revealing substantial variability that reflects marked disparities in the insertion windows across various VR weaknesses. A closer examination of the insertion windows for individual CWE types shows that VR weaknesses associated with \textit{Expression Issues} (CWE-569) and \textit{Type Errors} (CWE-136) tend to be introduced quickly. 
This is likely due to the fact that such VR weaknesses typically originate from simple coding mistakes or minor logic flaws.
In contrast, more complex VR weaknesses like \textit{User Interface Security Issues} (CWE-355), \textit{Signal Errors} (CWE-387) and \textit{Audit / Logging Errors} (CWE-1210) exhibit longer insertion windows, likely due to their systemic nature and the larger scope of changes required for their manifestation.

\insightbox{
There is substantial variability in the time it takes for VR weaknesses to be introduced, with some (e.g., CWE-569, CWE-136) appearing quickly as a result of simple coding oversights, whereas others (e.g., CWE-355, CWE-387, CWE-1210) take longer to surface due to their complex and embedded nature within the system.
}

\subsubsection{RQ5: How are VR weaknesses introduced into VR software?}
\label{sec:rq5_answer}
\ 
\newline
\indent 
To understand the root causes of VR weakness introduction, we analyze key contributing factors such as commit goal, commit frequency, the scale and frequency of code and file changes, and the adoption of third-party libraries.

We first extract 8,288 commit goals from 7,595 WCCs of 1,681 VR weaknesses, noting that a single WCC may correspond to multiple goals. Among these, the most prevalent goal is \textit{introduction of new features}, comprising 36.5\% of the total goals, followed by \textit{enhancements} at 24.4\%, \textit{refactoring} at 20.3\%, and \textit{bug fixing} at 18.8\%. The results suggest that adding new functionality is the most weakness-prone activity, likely due to the complexity of integration and the lack of thorough security consideration during the feature design phase. 

}

Subsequently, we investigate the relationship between commit frequency (\(F_{com}\)) and the introduction of VR weaknesses. The findings reveal no significant correlation, implying that the emergence of such weaknesses is more closely associated with the characteristics of code modifications or specific developer behaviors rather than the general level of repository activity.

\begin{figure}[!htbp]
    \centering
    \includegraphics[width=0.7\columnwidth]{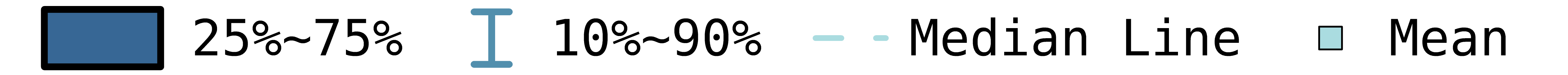}
    \includegraphics[width=0.9\columnwidth]{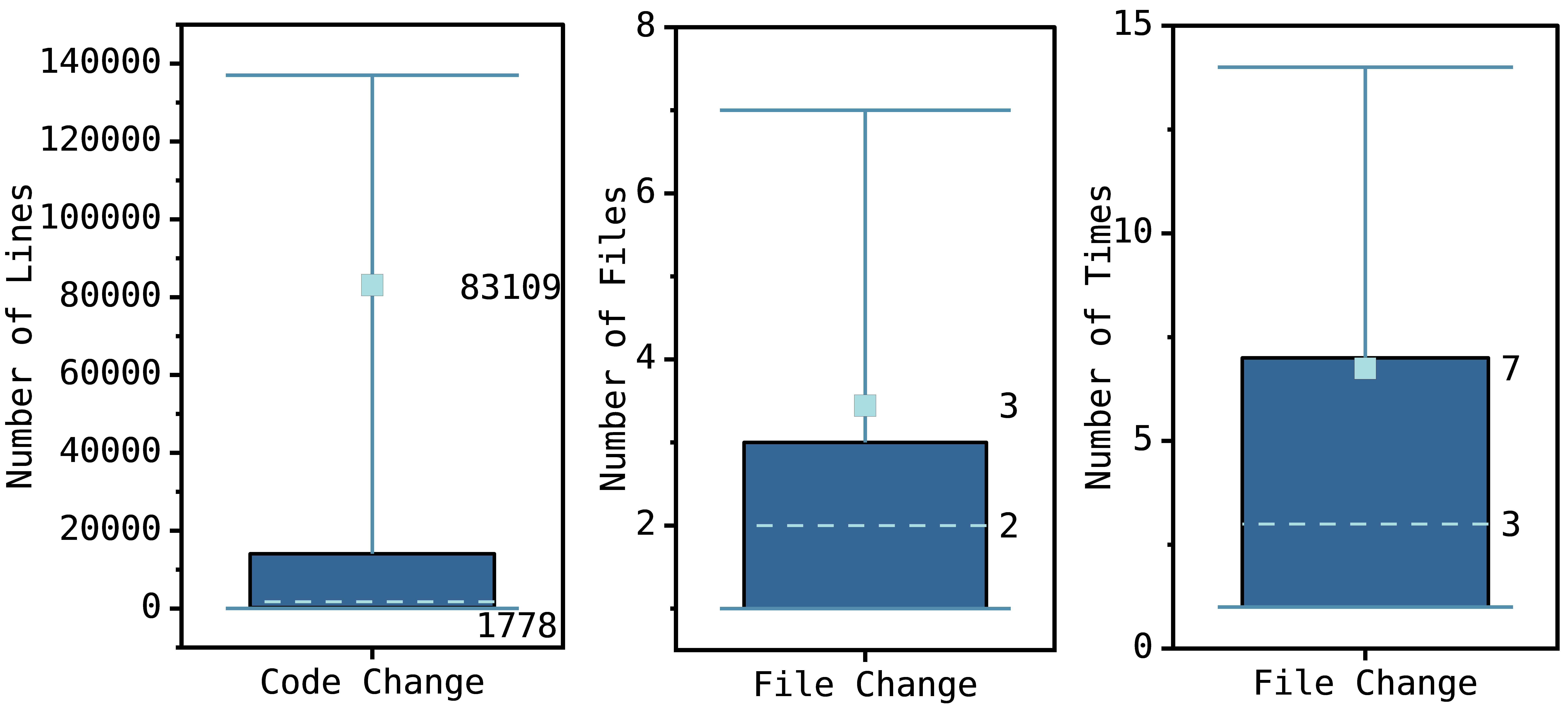}
    \caption{Impact of code and file edits on weakness introduction.}
    \label{fig:vul_intro_method}
    \vspace{-0.3cm}
\end{figure}

We further examine how code and file changes contribute to the introduction of VR weaknesses. {As shown in Figure \ref{fig:vul_intro_method}, the left box plot captures the scale of code edits during the insertion window \(t_{0,1}\), measured by the sum of lines added and deleted in the commits. The results reveal a broad range, with most of the VR weaknesses linked to relatively small-scale edits but others arising from extensive code modifications.} The middle box plot shows the number of files involved in each weakness, with a median of 2, an average of 3, and a maximum of 7 files. 
This suggests that while most weaknesses arise from localized edits, a subset reflects complex, system-wide changes likely driven by cross-component dependencies. The right box plot displays the modification frequency of weakness-associated files, revealing that these files typically undergo repeated changes, with a median of 3, an average of 7, and a maximum of 14 modifications. While most weakness-associated files undergo moderate changes, a subset experiences frequent modifications, indicating that high update frequency may be a risk factor for weakness introduction and should be a focus for proactive security monitoring.

\insightbox{
Most security weaknesses stem from localized, iterative small-scale modifications, highlighting the necessity of integrating lightweight and continuous security checks into daily development workflows, rather than relying solely on post hoc reviews.
}

We proceed to examine whether the identified VR weaknesses originate from third-party libraries and how their integration into VR projects affects the propagation of such VR weaknesses. We consider a set of 14 third-party libraries that are extensively utilized in current VR development, including Oculus, SteamVR, and the Google VR SDK, as listed in \cite{chen2025unveiling}. Table~\ref{tab:third-party} presents the vulnerable libraries, the names and types of projects depending on them, the number of affected files, the number of weaknesses introduced, and the number of unique first WCCs originating from each library. 
It is worth noting that first WCCs may be shared by multiple weaknesses either within a single project or across different projects that rely on the same library, reflecting the extent to which a library can serve as a distinct source of VR weaknesses.

\begin{table}[!htbp]
\centering
    \caption{Vulnerable Third-Party Libraries and Their Associated Impacts}
    \resizebox{\columnwidth}{!}{%
    \begin{tabular}{c|c|c|c|c|c}
        \hline
        \textbf{\begin{tabular}[c]{@{}c@{}}Library\end{tabular}} & \textbf{Project} & \textbf{Project class} & \textbf{\begin{tabular}[c]{@{}c@{}}Number of\\ affected files\end{tabular}} & \textbf{\begin{tabular}[c]{@{}c@{}}Number of\\ weaknesses\end{tabular}} & \textbf{\begin{tabular}[c]{@{}c@{}}Number of\\ first WCCs\end{tabular}} \\ \hline
        \multirow{2}{*}{\textbf{VRTK}} & VRTK & Development Tool & 27 & 22 & \multirow{2}{*}{23} \\ \cline{2-5}
         & VRTK-GearVR-Test & Development Tool & 27 & 22 &  \\ \hline
        \textbf{XR} & Technologies\_InputSystem & Development Tool & 14 & 4 & 4 \\ \hline
        \textbf{openvr} & MixedRealityToolkit-Unity & Development Tool & 4 & 4 & 3 \\ \hline
        \multirow{4}{*}{\textbf{steamvr}} & vhvr-mod & Development Tool & 1 & 1 & \multirow{4}{*}{4} \\ \cline{2-5}
         & VRTK-GearVR-Test & Development Tool & 1 & 1 &  \\ \cline{2-5}
         & ViveGrip & Development Tool & 1 & 2 &  \\ \cline{2-5}
         & nomai-vr & Application & 2 & 1 &  \\ \hline
    \end{tabular}
    }
    \label{tab:third-party}
    \vspace{-0.3cm}
\end{table}

We observe that VR weaknesses linked to these vulnerable libraries appeared in 7 projects overall, including 1 application and 6 development tools. Among the vulnerable libraries, \verb|VRTK| is the primary contributor to weakness introductions, being involved in {54 out of the 77 affected files}. This suggests that the complexity and ubiquitous integration of middleware libraries—such as those responsible for coordinating physics, input, and rendering subsystems—substantially expands the attack surface. Consequently, security review processes should pay special attention to widely used middleware libraries like \verb|VRTK|, and development workflows must incorporate exhaustive testing mechanisms when integrating diverse third-party components. However, it is worth noting that although libraries contribute to 34 first WCCs, this relatively small number suggests that most weaknesses stem from developer-written code rather than third-party libraries.

\insightbox{Third-party libraries constitute a source of VR weaknesses, with VRTK exhibiting the highest susceptibility, warranting rigorous and context-aware security audits of external dependencies. Nonetheless, the majority of weaknesses originate from developer-authored code rather than third-party dependencies.
}

\subsubsection{RQ6: How does developer status affect VR weaknesses introduction?}
\label{sec:rq6_answer}
\ 
\newline
\indent 
We explore the influence of developer pressure and expertise on the introduction of VR weaknesses by examining developer workload, measured by \(WL_{commit}\) and \(WL_{code}\), and developer experience, measured by \(Exp\). Figure \ref{fig:developers}\subref{fig:workload} illustrates the impact of workload on WCCs. We observe that developers with a high workload are responsible for the majority of WCCs, comprising 73.92\% for \(WL_{commit}\) and 69.23\% for \(WL_{code}\), significantly surpassing those with medium or low workloads. This implies that high-pressure environments, characterized by substantial code modifications or frequent commits, elevate the risk of introducing weaknesses. 
Figure \ref{fig:developers}\subref{fig:tenure} shows that expert developers contribute the largest proportion of WCCs, accounting for 65.53\%, overshadowing their less-experienced counterparts. This finding highlights a paradox: while experts are typically more capable and experienced, their extensive involvement in critical or complex system components may increase their likelihood of introducing VR weaknesses. Conversely, newcomers and medium-tenure developers tend to contribute fewer weaknesses, possibly owing to their restricted involvement in high-risk or mission-critical components.

\begin{figure}[!htbp]
\vspace{-0.5cm}
    \centering
    \subfloat[]{
        \includegraphics[width=0.52\columnwidth]{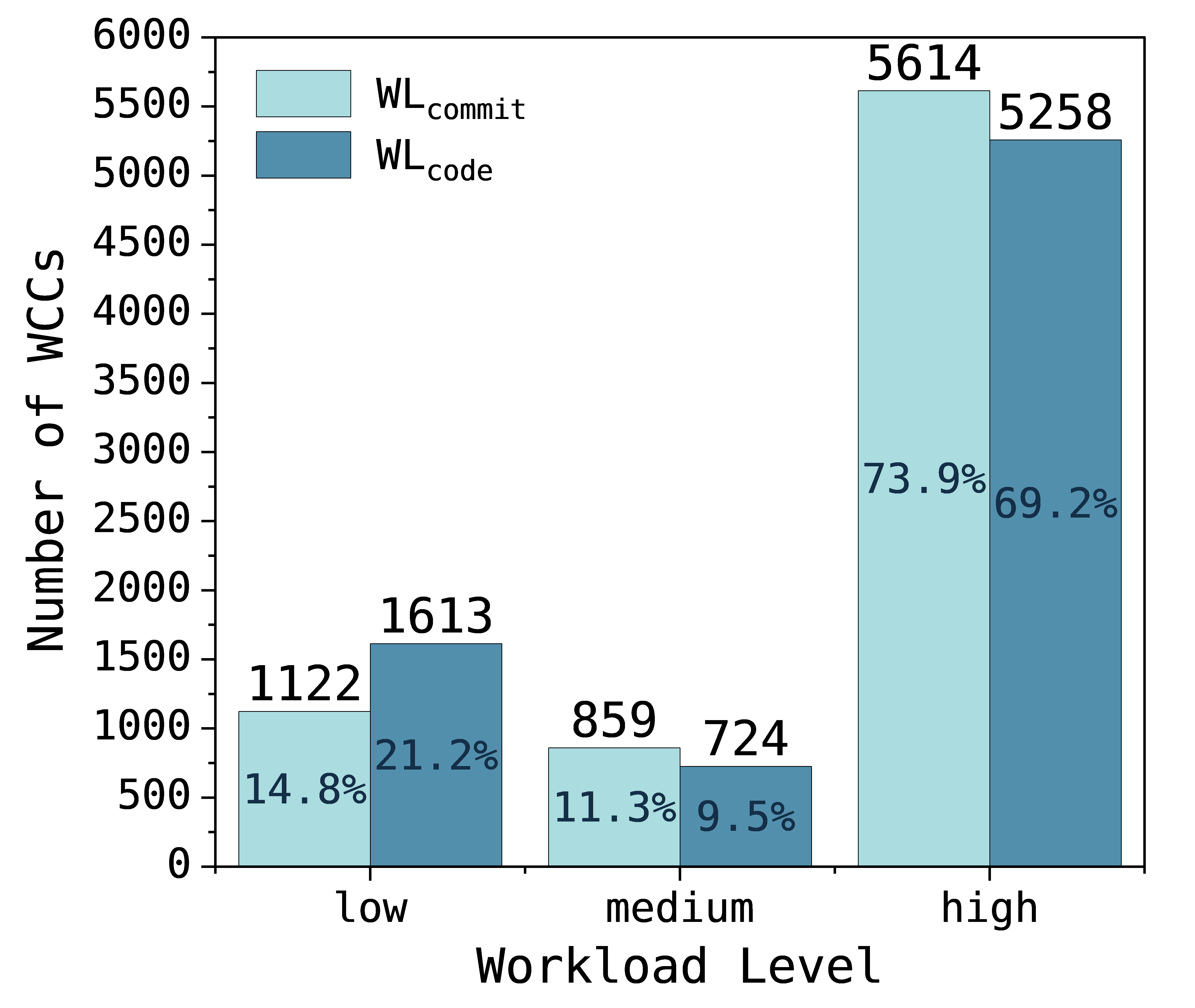}
        \label{fig:workload}
    }
    \hfill
    \subfloat[]{
        \includegraphics[width=0.38\columnwidth]{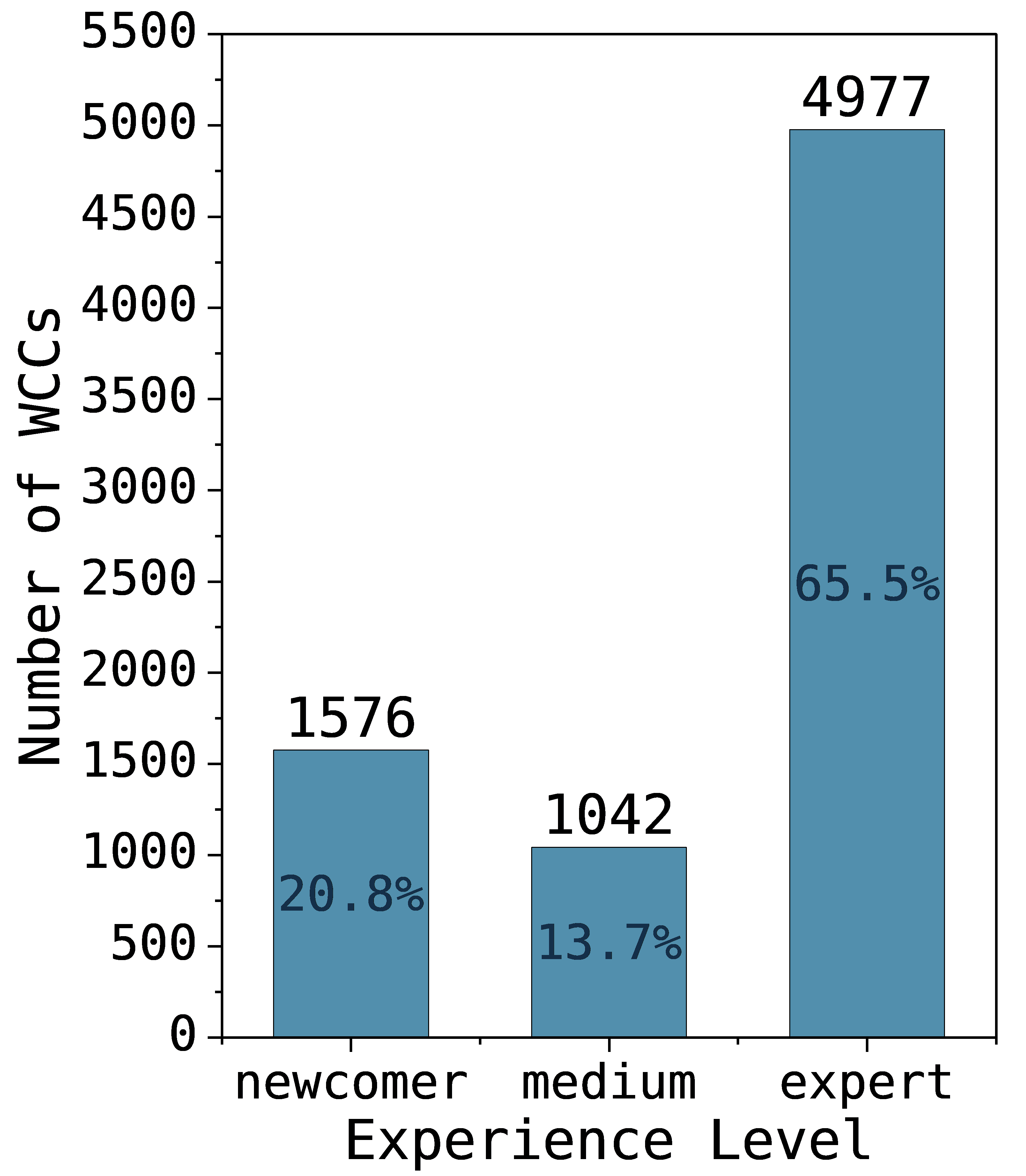}
        \label{fig:tenure}
    }
    \caption{The impact of developer status on WCCs: (a) workload (b) experience.}
    \label{fig:developers}
    \vspace{-0.5cm}
\end{figure}

\insightbox{
Heavy workloads amplify error rates, and experts—despite their proficiency—often introduce VR weaknesses due to their involvement in intricate, high-risk tasks.  This underscores the necessity of balanced workload distribution and reinforced safeguards for expert-generated code.}

\subsection{Security Weaknesses Survival Analysis}

\subsubsection{RQ7: How do the lifetimes of different VR weaknesses vary in diverse contexts?}
\label{sec:rq7_answer}
\ 
\newline
\indent 
To answer this RQ, we analyze the average lifetime (\(t_{1,3}\)) of VR weaknesses across various CWE types, as presented in Figure \ref{fig:life_time}, which reveals substantial disparities among them. Specifically, the left box displays the \(t_{1,3}\) distribution of all VR weaknesses. The median lifetime is relatively short, at approximately 4 days, indicating that a significant portion of VR weaknesses are fixed promptly after identification. However, the variability is notable, suggesting that certain VR weaknesses remain unresolved for extended periods. These VR weaknesses likely correspond to more intricate issues that require substantial investigation or structural changes in the project.

Additionally, distinct lifetime patterns are observed across various CWE types. Weaknesses with shorter lifespans typically correspond to easily detectable and remediable CWEs, including \textit{User Interface Security Issues} (CWE-355), \textit{String Errors} (CWE-133), \textit{Numeric Errors} (CWE-189) and \textit{Audit/Logging Errors} (CWE-1210).
These VR weaknesses often trigger immediate failures or errors, which are readily captured by standard testing and monitoring tools. In contrast, certain weaknesses such as \textit{Signal Errors} (CWE-387) and \textit{Type Errors} (CWE-136) exhibit longer lifespans. The persistence of \textit{Signal Errors} (CWE-387) is linked to the inherent complexity of hardware-software interaction protocols in VR systems, where intricate signaling mechanisms require cross-layer coordination for effective remediation. Similarly, \textit{Type Errors} (CWE-136) manifest through non-deterministic failure patterns that only emerge under specific runtime conditions, significantly delaying their detection and correction.

\begin{figure}[!ht]
    \centering
    \includegraphics[width=0.7\linewidth]{figs/legend2.png}
    \includegraphics[width=\linewidth]{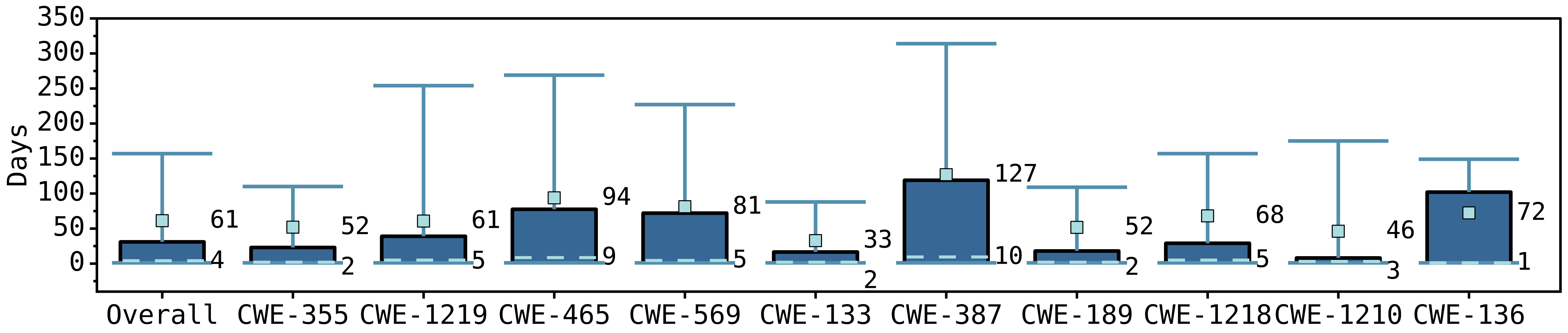}
    \caption{Lifetime of weaknesses across different CWE types.}
    \label{fig:life_time}
    \vspace{-0.5cm}
\end{figure}

\begin{figure}[!htbp]
    \centering
    \includegraphics[width=0.7\linewidth]{figs/legend2.png}
    \includegraphics[width=\linewidth]{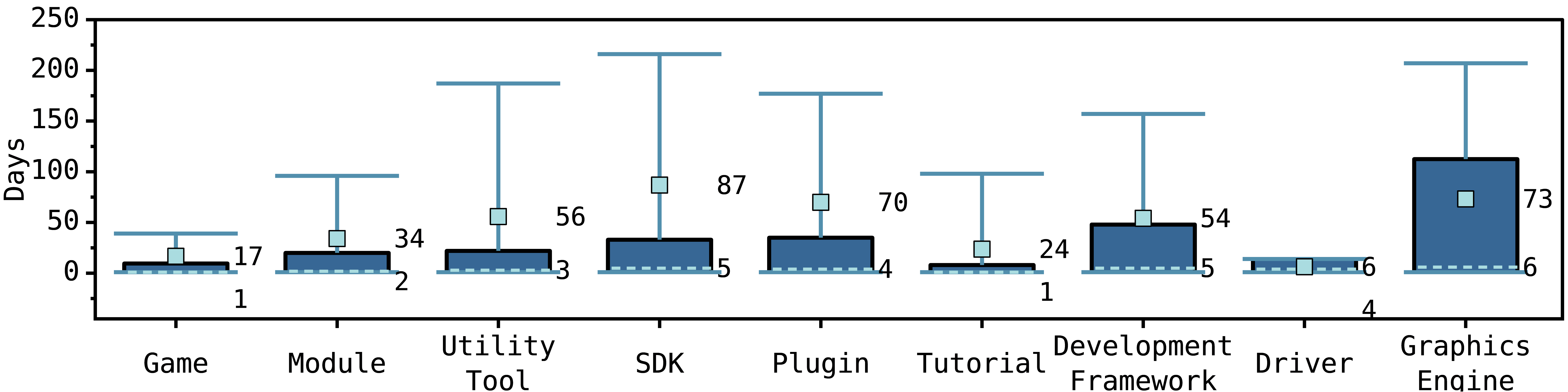}
    \caption{Lifetime of weaknesses across different project types.}
    \label{fig:life_time_types}
    \vspace{-0.3cm}
\end{figure}

We also analyze the lifetime of VR weaknesses with respect to different project types, as shown in Figure \ref{fig:life_time_types}. By referencing Table \ref{tab:project_type_cwe} and Figure \ref{fig:life_time}, we find that the results shown in Figure \ref{fig:life_time_types} are well-justified. Specifically, the short weakness lifetimes observed in \verb|game| and \verb|tutorial| projects can be attributed to the high presence of CWE-355, which accounts for 64.38\% and 88.89\% of the weaknesses in these projects, respectively, and is characterized by a relatively short persistence period. In contrast, \verb|graphics engines| projects exhibit considerably longer weakness lifetimes, primarily due to the 29.63\% prevalence of CWE-136, a type associated with extended weakness duration.

\begin{figure}[!htbp]
    \centering
    \includegraphics[width=0.7\linewidth]{figs/legend2.png}
    \begin{minipage}{0.46\columnwidth}  
        \centering
        \includegraphics[width=0.7\textwidth]{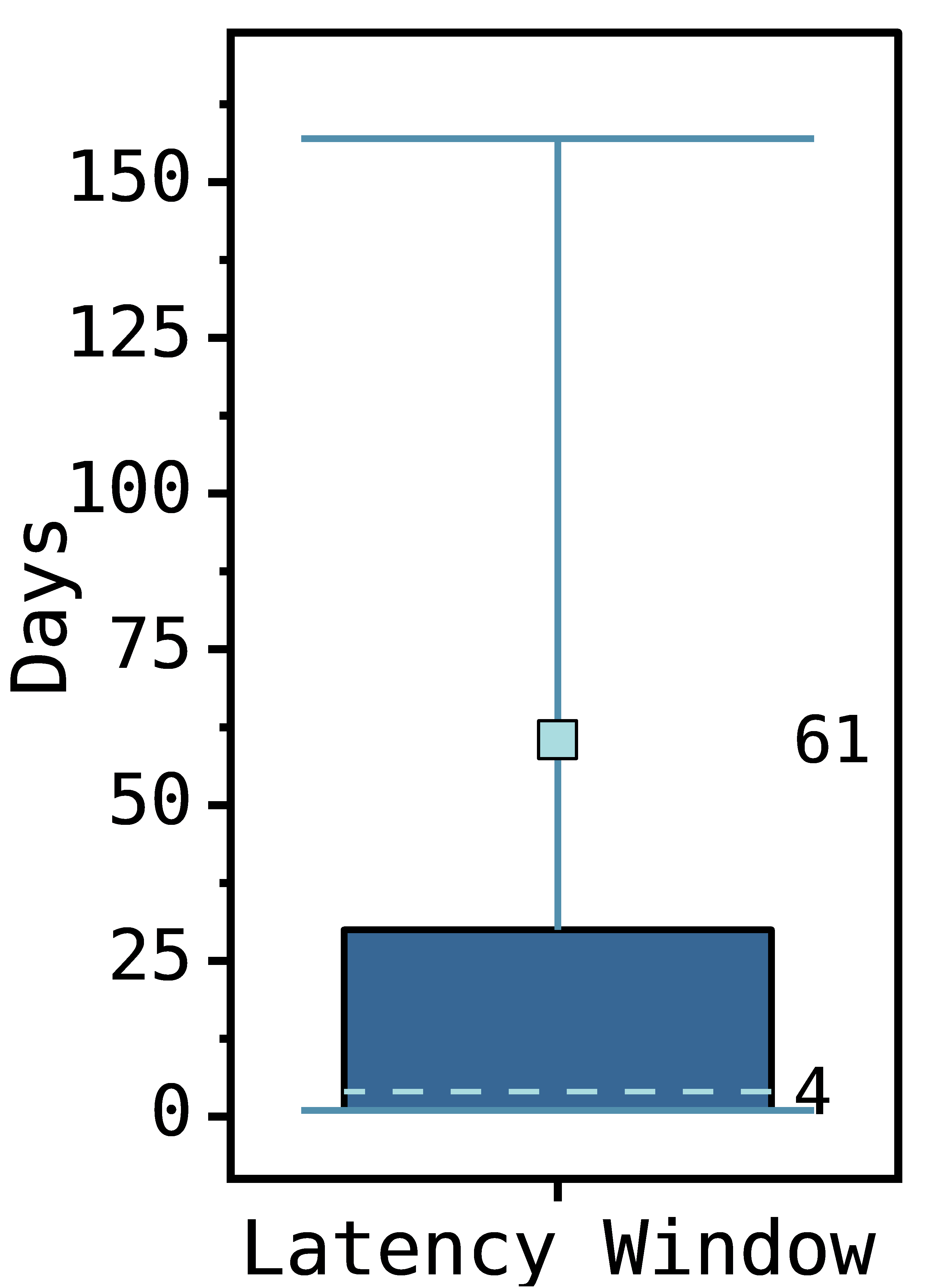}
        \caption{Latency window.}
        \label{fig:latency_window}
    \end{minipage}
    \hspace{0.01\columnwidth}
    \begin{minipage}{0.46\columnwidth}  
        \centering
        \includegraphics[width=0.72\textwidth]{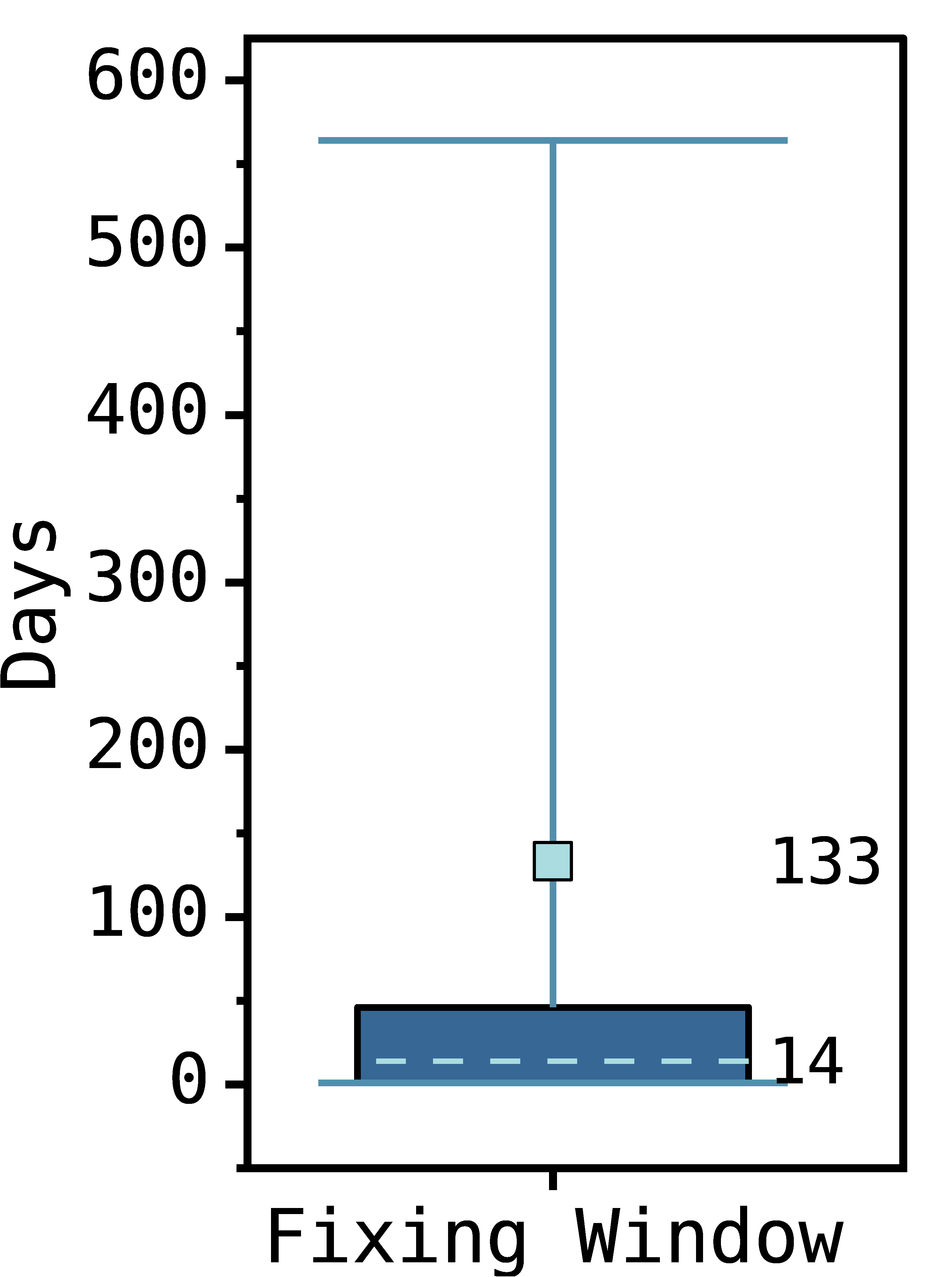}
        \caption{Fixing window.}
        \label{fig:fixing_window}
    \end{minipage}
    \vspace{-0.3cm}
\end{figure}

We further examine the latency window of VR weaknesses, defined as the duration during which a weakness exists in the system without being detected. As shown in Figure \ref{fig:latency_window}, the median latency window is 4 days after rounding, indicating that most VR weaknesses are detected and handled promptly. However, the substantial variability suggests that some weaknesses remain undetected for extended periods, thus increasing their likelihood of exploitation.

Lastly, we investigate the fixing window of VR weaknesses, defined as the duration between the first attempt to fix the weakness and its complete resolution. We begin by examining the number of times each VR weakness is fixed. It is found that 2 weaknesses are fixed three times, 23 weaknesses are fixed twice, and the rest 1,656 are fixed only once. The result reveals that over 98.5\% of the weaknesses are resolved with a single fix, with only 1.5\% requiring multiple iterations. Therefore, the average fixing window across all weaknesses is very short, which directly explains why the latency window in Figure \ref{fig:latency_window} closely approximates the overall lifetime in Figure \ref{fig:life_time} after rounding. Figure \ref{fig:fixing_window} shows the number of days required to fix the 25 VR weaknesses that needed more than one fix. It is observed that these weaknesses have substantially longer fixing windows, indicating a higher level of complexity in their resolution.

\insightbox{
Owing to prompt response and efficient remediation, the overall lifetime of VR weaknesses is notably short, with a median of approximately 4 days. Nevertheless, substantial disparities persist across different types of weaknesses and project categories.}

\begin{figure}[!htbp]
    \centering
    \includegraphics[width=0.7\linewidth]{figs/legend2.png}
    \includegraphics[width=0.85\linewidth]{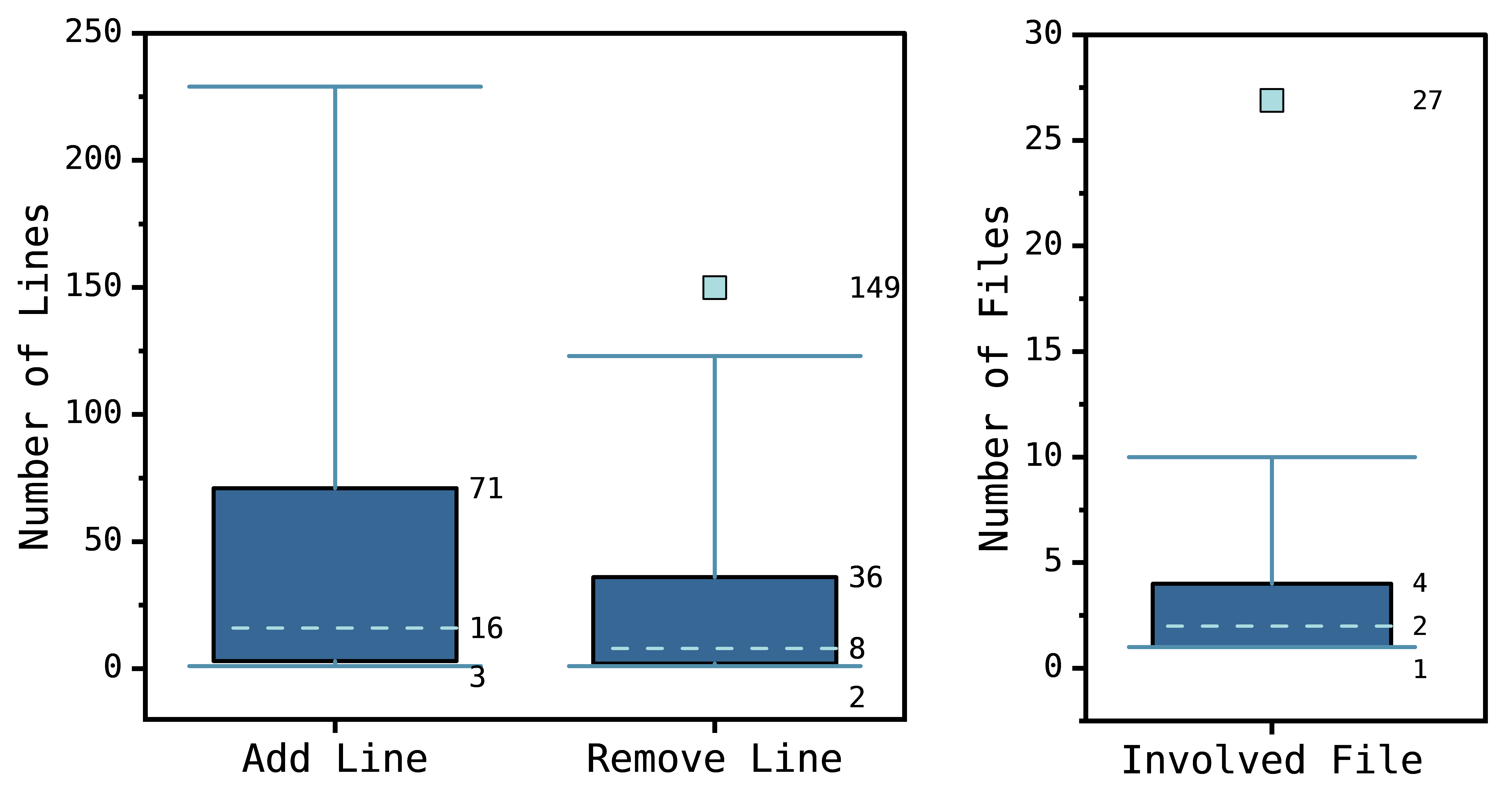}
    \caption{Methods and impacts of VR weakness remediation.}
    \label{fig:fix_method}
    \vspace{-0.2cm}
\end{figure}

\subsubsection{RQ8: How are VR weaknesses removed from the source code?}
\label{sec:rq8_answer}
\ 
\newline
\indent 
We investigate the nature of code changes and the number of files involved in fixing commits to identify common patterns in the resolution of VR weaknesses. From the perspective of code changes, the results show a balanced use of both additions and deletions in VR weaknesses fixes. Specifically, among the 1,708 total fixes, 98 (5.73\%) involve only additions to the code, and 81 (4.74\%) involve only deletions. The remaining 1,529 fixes (89.53\%) involve both operations, indicating that most fixes seek to refine existing functionality while integrating new logic to ensure security.

The left plot of Figure \ref{fig:fix_method} shows the number of lines added and removed during the fixing process. We observe that the majority of fixes are small-scale, involving fewer than 71 lines. However, a few outliers involve substantial code changes, likely due to complex security flaws or architectural deficiencies. The comparison between the two boxes also reveals a slight bias toward code addition, indicating a tendency to resolve weaknesses by introducing new logic.

The right plot of Figure \ref{fig:fix_method} shows the number of files involved in fixing commits. The results show that most fixes involve only one or two files, reflecting highly localized changes. This observation points to a key characteristic of VR software: its scene-based modularity. Each scene functions in relative isolation, with limited cross-file dependencies on others. As a result, most modifications remain localized within a specific scene.
Nevertheless, certain weaknesses have a system-wide impact, necessitating modifications across multiple files—{more than 10} in some cases. These outliers highlight that, while scene isolation simplifies maintenance, critical issues affecting shared modules or inter-scene logic can still necessitate system-wide changes.

\insightbox{Security weaknesses in VR software are typically addressed in a localized and efficient manner, with over 98\% resolved in a single attempt and more than 75\% requiring modifications to no more than two files. Developers tend to favor code addition over removal, reflecting a patch-based remediation approach possibly aimed at minimizing disruption to existing functionality.}

\subsubsection{RQ9: How effective are current code analysis tools in detecting VR weaknesses in VR software?}
\label{sec:rq9_answer}
\ 
\newline
\indent 
Dynamic analysis involves executing a program in a controlled environment and monitoring its behavior at runtime. While effective in many traditional software contexts, dynamic analysis becomes particularly challenging when applied to VR software due to several inherent constraints. (i) VR systems require complex runtime environments that incorporate real-time sensor inputs, motion tracking, and immersive graphical rendering. Constructing such environments for testing purposes is both costly and technically demanding. (ii) Dynamic analysis relies on well-designed test cases to explore diverse execution paths, but the interactive and unpredictable nature of VR applications makes it extremely difficult to achieve sufficient coverage. (iii) The instrumentation involved in dynamic analysis (e.g., memory monitoring, system call tracing) imposes heavy resource overhead, adversely affecting performance in latency-sensitive VR applications. Together, these limitations introduce considerable technical barriers to the  application of dynamic analysis in detecting VR weaknesses.

Compared to dynamic analysis, static analysis operates without the need to execute the program or reproduce the hardware and runtime environments, making it more suitable for VR software \cite{ernst2003static}.
Given that 88.3\% of the VR projects under study are implemented in C\#, and that most existing static analysis tools provide limited support for this language, we select CodeQL for evaluation, considering its powerful C\# support and active maintenance. Specifically, we adopt the official CodeQL query suite, \verb|CodeQL-main|, which includes a collection of default queries maintained by GitHub Advanced Security \footnote{\url{https://github.com/codeql}} for detecting common security issues. Despite its utility in identifying C\# VR weaknesses, this suite does not achieve full coverage of the top 20 CWE types commonly observed in VR software. The intersection contains six weakness types, as shown in Figure~\ref{fig:venn}. 

\begin{figure}[!htbp]
    \centering
    \includegraphics[width=0.8\linewidth]{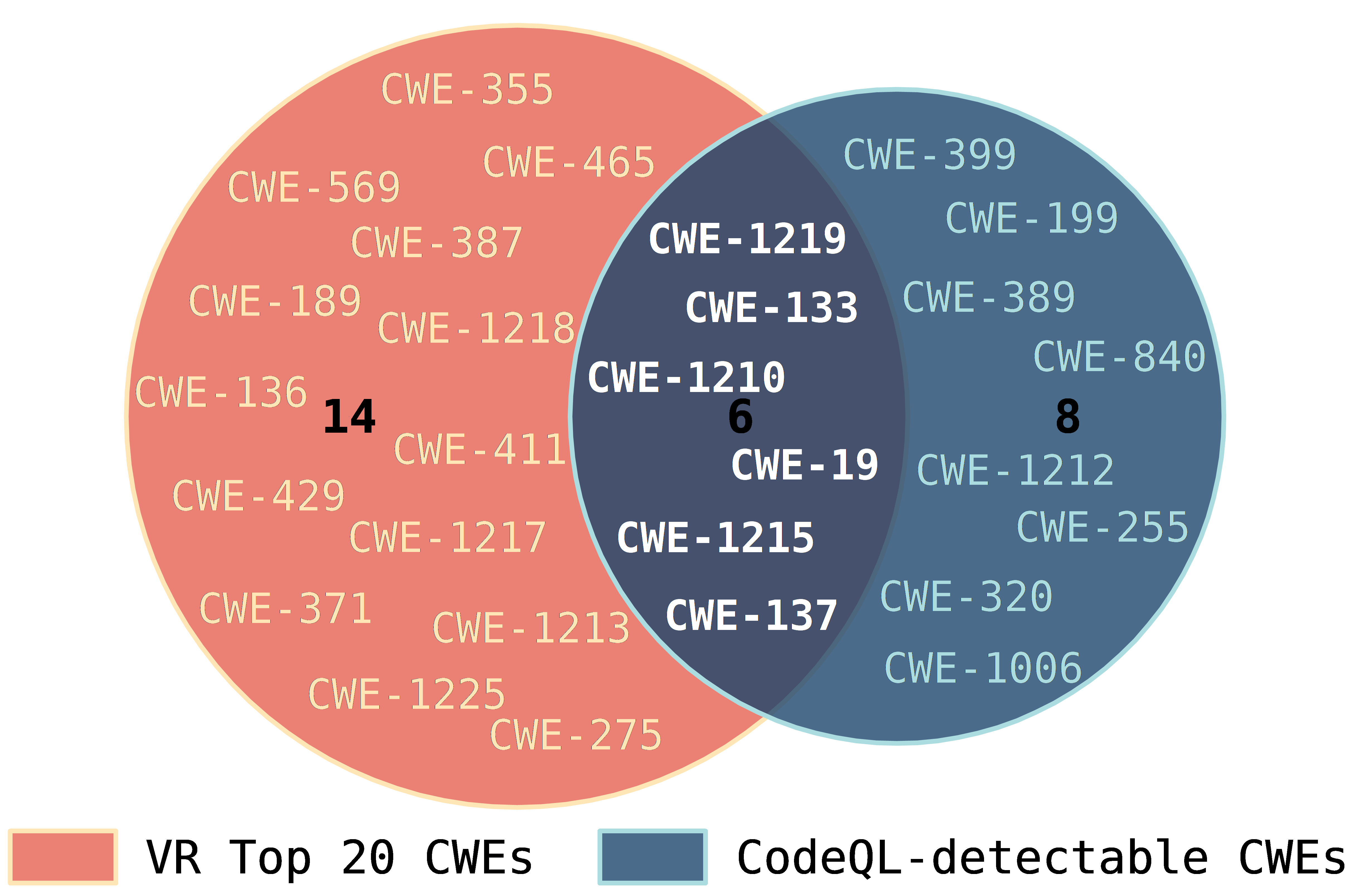}
    \caption{VR top 20 CWEs vs. CodeQL-detectable CWEs.}
    \label{fig:venn}
    \vspace{-0.3cm}
\end{figure}

{The process of preparing Unity-based C\# projects for CodeQL static analysis is non-trivial, involving (i) generating \verb|.sln| and \verb|.csproj| files via the Unity engine, (ii) resolving all dependencies through the \verb|dotnet| build process to enable CodeQL database creation, and (iii) executing queries under the constraint that no more than ten can be run concurrently. Therefore, we select six representative projects with confirmed \textit{File Handling Issues} (CWE-1219) as case studies, recognizing this as the most prevalent weakness type detectable via CodeQL. The project information and corresponding analysis results are summarized in Table~\ref{tab:codeql_detect}.

\begin{table}[!htbp]
    \centering
    \caption{Overview of Projects Analyzed with CodeQL and Their CWE-1219 Detection Results} 
    \label{tab:codeql_detect}
    \resizebox{\columnwidth}{!}{
    \begin{tabular}{c|c|>{\centering\arraybackslash}p{0.4\columnwidth}|c}
         \hline
         \textbf{Project name} & \textbf{Commit hash} & \textbf{Commit message description} & \textbf{Detection result} \\
         \hline
         AdGoBye & e83b338 & Fix an error accessing file ``\_data" that does not exist. & None \\ 
         \hline
         Creator & ab81c26 & Fix an error in using hard-code file path. & None \\ 
         \hline                                              
         VRCToolBox & 233571d & Add check file exists. & None  \\ 
         \hline
         UnityPlugin & f729b65 & Fix assets serialization and path resolution issues for folder ``StreamingAssets". & None \\ 
         \hline 
         UMI3D-SDK & 923d355 & Add check file exists. & None \\ 
         \hline 
         InputSystem & 8becec7 & Add check file exists. & None \\ 
         \hline 
    \end{tabular}
    }
    \vspace{-0.3cm}
\end{table}

Unexpectedly, CodeQL fails to detect any of the identified weaknesses. This limitation likely stems from several factors: (i) CodeQL implements only a narrow interpretation of CWE-1219, specifically focusing on \textit{Path Traversal} (CWE-22) that involves insecure handling of user-controlled file paths. The other subcategories of CWE-1219, such as missing file existence checks, insecure use of asset folders, and improper file resolution, are not included in the default query set. (ii) Many file-handling operations in VR software employ framework-specific APIs such as Unity's \verb|Resources.Load| or access specific folders like \verb|StreamingAssets|. These implementation patterns lie beyond CodeQL's detection scope since its generic C\# analysis rules primarily focus on standard .NET I/O libraries, consequently omitting these APIs from both data flow and taint analysis models. (iii) VR software frequently involves complex control flows, engine-specific abstractions, and data flows that span across non-code assets, all of which are not adequately addressed by current static analysis tools.
These constraints expose a fundamental gap in current static analysis techniques for VR software, emphasizing the necessity of developing specialized detection rules that account for the unique architectural and interaction characteristics of VR systems.

\insightbox{Traditional static and dynamic analysis methods may be less effective at detecting VR weaknesses due to the unique characteristics of VR software, such as complex interactions, framework-specific APIs, and heavy reliance on non-code assets—underscoring the potential need for detection approaches tailored to VR's distinct architecture and runtime behavior.
}

\vspace{-0.5cm}
\section{Discussion}
\label{sec:discussion}

\subsection{Recommendations for Advancing Research and Development}
\subsubsection{For Researchers}
\ 
\newline
\indent 
This study opens several avenues for future research on the identification and prevention of VR weaknesses in VR software. The observed trends in weakness types—such as the persistence of user interface errors, resource management flaws, and authorization issues—warrant deeper investigation into their underlying causes. Future work could examine the interplay between development practices, project characteristics, and resulting security outcomes. Another important finding is the frequent occurrence of VR weaknesses during the file creation phase. This highlights the need for research on predictive models and early-warning systems that can identify potential weaknesses before they are introduced or propagated throughout the development lifecycle.

It is also important to study more flexible automated VR weaknesses analysis tools for VR software. Current static analysis tools provide limited support for C\# and VR weakness types, which highlights the need to design VR weakness query rules.
Besides, researchers are encouraged to explore hybrid approaches to better capture control flows among users, hardware, code, and assets. In particular, symbolic execution can be a promising direction. It enables path-sensitive analysis without requiring concrete inputs or full runtime environments, making it suitable for analyzing the complex user interactions in VR software.

\subsubsection{For Tool Developers}
\ 
\newline
\indent 
VR software is typically composed of tightly coupled files and components, where changes in one module can have far-reaching effects on others. In this context, system-wide dependency visualization tools are essential for revealing these interconnections. Such tools can assist developers in tracing potential security risks across the entire system, promoting holistic security assessments rather than addressing weaknesses in isolation.

Maintaining synchronization between 3D modeling tools and development engines is also crucial, as discrepancies arising from version mismatches can introduce errors during the export and import of models. Tools that automate compatibility checks and assist with format conversion can significantly reduce integration challenges, thereby minimizing unintended defects and inconsistencies in VR applications. Additionally, implementing automated version control for assets and configurations can streamline collaboration and mitigate risks associated with file overwrites and dependency errors.

\subsubsection{For Project Developers and Teams}
\ 
\newline
\indent 
Developers should adopt security-first practices across all stages of the software development lifecycle. For VR application projects, robust file management and data validation mechanisms are essential to meet real-time performance requirements while preserving system integrity. In the context of development tools, minimizing memory-related errors and computational logic issues are critical for enhancing usability and reducing weaknesses. Importantly, early-stage security assessments, especially during file creation, play a critical role in preventing downstream weaknesses.

Project teams should implement continuous security assessments throughout both development and maintenance phases. These assessments must account for both localized and system-wide changes to effectively identify and mitigate potential weaknesses. Close collaboration between security specialists and developers is crucial to ensure that iterative updates undergo thorough security reviews, thereby minimizing the risk of introducing new weaknesses. Additionally, proper workload distribution among team members can further contribute to reducing the occurrence of weaknesses.

\vspace{-0.4cm}
\subsection{Threats to Validity}
\label{sec:threats}

\subsubsection{Internal Validity}
\begin{itemize}
    \item Our approach relies on commit message analysis and code changes to identify security-related commits, which may misidentify weaknesses. To mitigate this, we employ multiple sentence-to-vector models requiring consensus among at least four models, reducing false positives. Our manual verification of 318 samples confirms the reliability of the obtained VR weaknesses. The SZZ algorithm has limitations when dealing with complex refactoring; we address this through enhanced file filtering and deduplication.
    \item Our VR weakness definition is based on CWE-699, which may not encompass all VR-specific security issues. We rely on developers accurately documenting security fixes, which isn't always the case. The lifecycle attributes defined in our study may be affected by varying development practices across projects.
    \item We employ appropriate statistical methods and clearly state confidence levels for statistical inferences. For qualitative assessments, multiple researchers independently review the data. However, our snapshot analysis may not capture long-term security practice trends in these evolving open-source projects.
\end{itemize}

\subsubsection{External Validity}
\ 
\newline
\indent 
Our dataset comprises 334 open-source VR projects, predominantly in C\# (88.3\%), reflecting the current VR development landscape. Findings may not generalize to projects in other languages, closed-source applications, or those hosted on platforms other than GitHub.

\vspace{-0.35cm}
\subsection{Applicability of the Findings}

While our study focuses on open-source VR software, the applicability of these findings to commercial VR software remains uncertain and is contingent upon various factors.

On one hand, both open-source and commercial VR software share common architectural principles, including scene-oriented structures, real-time interactions between multiple sources, and stringent performance demands. Similar to open-source VR software, commercial VR software also relies heavily on game engines like Unity or Unreal, and utilizes widely-used third-party libraries like VRTK, SteamVR and openVR. Therefore, insights gleaned from open-source VR, such as the predominance of user interface weaknesses, the early-stage introduction of weaknesses during development, and the presence of risks associated with third-party dependencies, are likely to be reflected in commercial VR systems as well.

On the other hand, commercial VR software may exhibit notable differences in several critical aspects. First, commercial software typically adheres to more stringent development lifecycles, which may help mitigate the prevalence of certain types of weaknesses. Second, commercial software tends to be larger in scale and more tightly integrated with proprietary services. While this integration offers advanced capabilities, it may create new attack vectors within proprietary services, an issue that is less common in open-source VR software, as it typically does not rely on proprietary dependencies. Finally, the higher level of professionalism among commercial software developers may influence the distribution and nature of weaknesses within the VR system.

Considering both the shared characteristics and distinctions, we believe that insights from open-source VR software can help inform the security improvements of commercial VR systems. However, effectively applying these insights hinges on a deeper understanding of the commercial development process, an area that still requires further exploration.

\section{Related Work}
\label{sec:related_work}

\subsection{Empirical Study of Software Vulnerability}

Empirical studies on the lifecycle of vulnerabilities in software have been extensively conducted to understand how vulnerabilities are introduced, persist, and are eventually resolved.
Altinkemer et al. \cite{Altinkemer2008Vulnerabilitiesa} analyze vulnerability patch lifecycles, highlighting resolution time variations based on complexity. Bosu et al. \cite{Bosu2014Identifyinga} examine the relationship between code changes and vulnerability introduction. 
Di Penta et al. \cite{DiPenta2009lifea} explore the evolution of static vulnerabilities in network systems. Zhang et al. \cite{Zhang2011Empiricala} focus on predicting the timing of future vulnerabilities. Tufano et al. \cite{Tufano2017When} investigate how vulnerabilities are introduced. Sliwerski et al. \cite{sliwerski2005changes} study the methods for fixing vulnerabilities.

While earlier empirical studies often targeted isolated phases, recent investigations have taken a broader view, systematically analyzing the complete vulnerability lifecycle \cite{Iannone2023Secret, Li2017Large-Scale, Shahzad2020Large, Liu2020large-scalea, ShiDoes}. 
For example, Iannone et al. \cite{Iannone2023Secret} analyze large-scale systems and find that vulnerabilities are predominantly introduced during feature development, with common patterns including delayed detection and patching. 
Shahzad et al. \cite{Shahzad2020Large} examine the characteristics of 56,077 leaked vulnerabilities over 15 years, identifying common issues across diverse products and evaluating containment and patching practices.

While prior research largely relies on publicly disclosed vulnerabilities from sources like the NVD, the novelty and unstandardized practices of VR development have led to limited disclosure. As a result, empirical studies on VR-related vulnerabilities remain scarce. 

\vspace{-0.4cm}
\subsection{Empirical Study of VR Software}
The empirical study of VR software has gained increased attention in recent years, though much of the existing research has focused on non-vulnerability aspects, such as developer practices, software performance, and user experiences. For instance, Oyelere et al. \cite{Oyelere2020Exploringa} investigate the development and usage patterns of VR applications in educational contexts. Grudzewski et al. \cite{Grudzewski2018Virtuala} explore the use of VR in marketing communication, analyzing its impact on message delivery, technology adoption, and user perception. Epp et al. \cite{Epp2021empirical} examine user experiences and operational performance in VR applications by analyzing complaints and trends in popular VR games. Huang et al. \cite{Huang2024Study} investigate code clones in VR open-source software, focusing on patterns of reuse and their implications for software maintenance. 

Recent efforts have begun to explore security weaknesses in open-source VR systems, particularly in VR applications. Rodriguez and Wang \cite{Rodriguez2017Empirical} pioneer this direction by identifying security issues such as frequent mis-commits of automatically generated files. Rzig et al. \cite{Rzig2023Virtual} conduct a case study on Unity-based VR software, proposing automated test case generation and quality assessment methods to enhance system security and reliability. Other studies have focused on specific weakness types and challenges in VR systems. Dastgerdy et al. \cite{Dastgerdy2024Virtual} examine security weaknesses in VR devices, emphasizing challenges such as securing real-time data, motion tracking, and other interactive components. Guo et al. \cite{Guo2025Empirical} investigate security and privacy issues in Oculus VR applications, analyzing vulnerabilities and security weaknesses related to user authentication, session management, and data privacy. 

While these studies shed light on prevalent weakness types and platform-specific security issues in VR systems, they do not examine the full lifecycle of weaknesses—how they emerge, persist, and are resolved. This reveals a significant gap in the current understanding of weakness evolution in VR software.

\vspace{-0.2cm}
\subsection{Vulnerability Detection and Tracing}
Vulnerability detection and tracing are fundamental aspects of software security research. Traditional approaches typically rely on structured databases of publicly disclosed vulnerabilities \cite{Iannone2023Secret, ShiDoes, Zhang2011Empiricala}, offering a comprehensive view of known issues. However, these methods are inherently limited to publicly available data and may not capture the full range of vulnerabilities present in software projects.

Lenarduzzi et al. \cite{Lenarduzzi2020OpenSZZ} and Borg et al. \cite{Borg2019SZZ} leverage issue tracking systems such as Jira \cite{Jira} and BugZilla \cite{Bugzilla} to extract vulnerability information, which is then linked to version control data using tools like Git. However, this approach requires access to specific issue tracking tools and the use of custom query languages, which restricts its generalizability across diverse software projects.

This study introduces a novel framework for detecting and tracing VR weaknesses using only GitHub commit data. By analyzing repository content, commit messages, and file changes, our approach bypasses the need for external databases or issue tracking systems—making it especially suitable for open-source VR projects.

\vspace{-0.3cm}
\section{Conclusion}
\label{sec:conclusion}

We have presented an empirical study of 334 open-source VR projects, uncovering key patterns in the evolution, introduction, and remediation of 1,681 VR weaknesses. To the best of our knowledge, this is the first study that systematically investigates VR weaknesses. We propose a novel framework that exclusively leverages GitHub commit data for the detection and tracking of VR weaknesses, and we construct the first benchmark specifically designed to evaluate such weaknesses in VR software. Guided by nine carefully designed research questions, our empirical analysis has yielded a number of actionable insights for developers, tool designers, and researchers. We also highlight several promising directions for future research, explore potential threats to validity, and discuss the relevance of our findings to commercial VR software.

\vspace{-0.3cm}
\section{Acknowledgment}
\label{sec:acknowledgment}

The authors would like to thank Prof. Kai Chen from the Institute of Information Engineering, Chinese Academy of Sciences, for his illuminating discussions and insightful comments on the manuscript, which provided significant guidance in the completion of this paper.

\vspace{-0.3cm}

\begin{thebibliography}{10}
\providecommand{\url}[1]{#1}
\csname url@samestyle\endcsname
\providecommand{\newblock}{\relax}
\providecommand{\bibinfo}[2]{#2}
\providecommand{\BIBentrySTDinterwordspacing}{\spaceskip=0pt\relax}
\providecommand{\BIBentryALTinterwordstretchfactor}{4}
\providecommand{\BIBentryALTinterwordspacing}{\spaceskip=\fontdimen2\font plus
\BIBentryALTinterwordstretchfactor\fontdimen3\font minus \fontdimen4\font\relax}
\providecommand{\BIBforeignlanguage}[2]{{%
\expandafter\ifx\csname l@#1\endcsname\relax
\typeout{** WARNING: IEEEtran.bst: No hyphenation pattern has been}%
\typeout{** loaded for the language `#1'. Using the pattern for}%
\typeout{** the default language instead.}%
\else
\language=\csname l@#1\endcsname
\fi
#2}}
\providecommand{\BIBdecl}{\relax}
\BIBdecl

\bibitem{GobbettiVirtual}
E.~Gobbetti and R.~Scateni, ``Virtual reality: Past, present and future,'' \emph{Studies in health technology and informatics}, vol.~58, pp. 3--20, 02 1998.

\bibitem{Smutny2023Learning}
P.~Smutny, ``\BIBforeignlanguage{en}{Learning with virtual reality: A market analysis of educational and training applications},'' \emph{\BIBforeignlanguage{en}{Interactive Learning Environments}}, vol.~31, no.~10, pp. 6133--6146, 2023.

\bibitem{Creed2024Inclusive}
C.~Creed, M.~{Al-Kalbani}, A.~Theil, S.~Sarcar, and I.~Williams, ``\BIBforeignlanguage{en}{Inclusive {{Augmented}} and {{Virtual Reality}}: {{A Research Agenda}}},'' \emph{\BIBforeignlanguage{en}{International Journal of Human--Computer Interaction}}, vol.~40, no.~20, pp. 6200--6219, 2024.

\bibitem{Javaid2020Virtual}
M.~Javaid and A.~Haleem, ``\BIBforeignlanguage{en}{Virtual reality applications toward medical field},'' \emph{\BIBforeignlanguage{en}{Clinical Epidemiology and Global Health}}, vol.~8, no.~2, pp. 600--605, 2020.

\bibitem{Akbulut2018effectiveness}
A.~Akbulut, C.~Catal, and B.~Y{\i}ld{\i}z, ``\BIBforeignlanguage{en}{On the effectiveness of virtual reality in the education of software engineering},'' \emph{\BIBforeignlanguage{en}{Computer Applications in Engineering Education}}, vol.~26, no.~4, pp. 918--927, 2018.

\bibitem{Martin-Gutierrez2017Virtual}
J.~{Mart{\'i}n-Guti{\'e}rrez}, C.~E. Mora, B.~{A{\~n}orbe-D{\'i}az}, and A.~{Gonz{\'a}lez-Marrero}, ``\BIBforeignlanguage{en}{Virtual {{Technologies Trends}} in {{Education}}},'' \emph{\BIBforeignlanguage{en}{EURASIA Journal of Mathematics, Science and Technology Education}}, vol.~13, no.~2, 2017.

\bibitem{Weiss1998Virtual}
P.~L. Weiss and A.~S. Jessel, ``\BIBforeignlanguage{en}{Virtual reality applications to work},'' \emph{\BIBforeignlanguage{en}{Work}}, vol.~11, no.~3, pp. 277--293, 1998.

\bibitem{Pujiono2024Augmented}
I.~P. Pujiono, A.~Asfahani, and A.~Rachman, ``\BIBforeignlanguage{en}{Augmented {{Reality}} ({{AR}}) and {{Virtual Reality}} ({{VR}}): {{Recent Developments}} and {{Applications}} in {{Various Industries}}},'' \emph{\BIBforeignlanguage{en}{Innovative: Journal Of Social Science Research}}, vol.~4, no.~4, pp. 1679--1690, 2024.

\bibitem{Berg2017Industry}
L.~P. Berg and J.~M. Vance, ``\BIBforeignlanguage{en}{Industry use of virtual reality in product design and manufacturing: A survey},'' \emph{\BIBforeignlanguage{en}{Virtual Reality}}, vol.~21, no.~1, pp. 1--17, 2017.

\bibitem{VirtualR14:online}
J.~Katatikarn, ``Virtual reality statistics: The ultimate list in 2024,'' \url{https://academyofanimatedart.com/virtual-reality-statistics/}.

\bibitem{Dastgerdy2024Virtual}
S.~Dastgerdy, ``Virtual {{Reality}} and {{Augmented Reality Security}}: {{A Reconnaissance}} and {{Vulnerability Assessment Approach}},'' 2024.

\bibitem{Vondracek2023Rise}
M.~Vondr\'{a}\v{c}ek, I.~Baggili, P.~Casey, and M.~Mekni, ``Rise of the metaverse’s immersive virtual reality malware and the man-in-the-room attack \& defenses,'' \emph{Comput. Secur.}, vol. 127, no.~C, Apr. 2023.

\bibitem{Silva2023Survey}
T.~Silva, S.~Paiva, P.~Pinto, and A.~Pinto, ``A survey and risk assessment on virtual and augmented reality cyberattacks,'' in \emph{2023 30th International Conference on Systems, Signals and Image Processing (IWSSIP)}, 2023, pp. 1--5.

\bibitem{Elliott2015Virtual}
A.~Elliott, B.~Peiris, and C.~Parnin, ``\BIBforeignlanguage{en}{Virtual {{Reality}} in {{Software Engineering}}: {{Affordances}}, {{Applications}}, and {{Challenges}}},'' in \emph{\BIBforeignlanguage{en}{2015 {{IEEE}}/{{ACM}} 37th {{IEEE International Conference}} on {{Software Engineering}}}}.\hskip 1em plus 0.5em minus 0.4em\relax Florence, Italy: IEEE, 2015, pp. 547--550.

\bibitem{Guo2025Empirical}
H.~Guo, H.-N. Dai, X.~Luo, G.~Xu, F.~He, and Z.~Zheng, ``An empirical study on meta virtual reality applications: Security and privacy perspectives,'' \emph{IEEE Transactions on Software Engineering}, vol.~51, no.~5, pp. 1437--1454, 2025.

\bibitem{Rodriguez2017Empirical}
I.~Rodriguez and X.~Wang, ``An empirical study of open source virtual reality software projects,'' in \emph{2017 ACM/IEEE International Symposium on Empirical Software Engineering and Measurement (ESEM)}, 2017, pp. 474--475.

\bibitem{Rzig2023Virtual}
D.~E. Rzig, N.~Iqbal, I.~Attisano, X.~Qin, and F.~Hassan, ``Virtual reality (vr) automated testing in the wild: A case study on unity-based vr applications,'' in \emph{Proceedings of the 32nd ACM SIGSOFT International Symposium on Software Testing and Analysis}, 2023, p. 1269–1281.

\bibitem{chandrashekar2023design}
N.~D. Chandrashekar, K.~King, D.~Gra{\v{c}}anin, and M.~Azab, ``Design \& development of virtual reality empowered cyber-security training testbed for iot systems,'' in \emph{2023 3rd Intelligent Cybersecurity Conference (ICSC)}.\hskip 1em plus 0.5em minus 0.4em\relax IEEE, 2023, pp. 86--94.

\bibitem{siddiqi2023secure}
S.~J. Siddiqi, M.~A. Jan, A.~M. Basalamah, and M.~Tariq, ``Secure teleoperated vehicles in augmented reality of things: A multichain and digital twin approach,'' \emph{IEEE Transactions on Consumer Electronics}, vol.~70, no.~1, pp. 956--965, 2023.

\bibitem{qamar2023systematic}
S.~Qamar, Z.~Anwar, and M.~Afzal, ``A systematic threat analysis and defense strategies for the metaverse and extended reality systems,'' \emph{Computers \& Security}, vol. 128, p. 103127, 2023.

\bibitem{nnamonu2023metaverse}
O.~Nnamonu, M.~Hammoudeh, and T.~Dargahi, ``{ Metaverse Cybersecurity Threats and Risks Analysis: The case of Virtual Reality Towards Security Testing and Guidance Framework },'' in \emph{2023 IEEE International Conference on Metaverse Computing, Networking and Applications (MetaCom)}.\hskip 1em plus 0.5em minus 0.4em\relax IEEE Computer Society, Jun. 2023, pp. 94--98.

\bibitem{NVD}
R.~Byers, C.~Turner, and T.~Brewer, ``\BIBforeignlanguage{en}{National {{Vulnerability Database}}}.''

\bibitem{VulDB}
``Vuldb vulnerability and threat intelligence | splunkbase,'' \url{https://splunkbase.splunk.com/app/4190}, 2024, accessed: 2024-12-31.

\bibitem{How-2025-03-17}
``How does virtual reality work? | coursera,'' \url{https://www.coursera.org/articles/how-does-virtual-reality-work}, 2025, accessed: 2025-03-17.

\bibitem{CWE}
\BIBentryALTinterwordspacing
``{{CWE}} - {{CWE-699}}: {{Software Development}} (4.15).'' [Online]. Available: \url{https://cwe.mitre.org/data/definitions/699.html}
\BIBentrySTDinterwordspacing

\bibitem{meneely2013patch}
A.~Meneely, H.~Srinivasan, A.~Musa, A.~R. Tejeda, M.~Mokary, and B.~Spates, ``When a patch goes bad: Exploring the properties of vulnerability-contributing commits,'' in \emph{2013 ACM/IEEE International Symposium on Empirical Software Engineering and Measurement}.\hskip 1em plus 0.5em minus 0.4em\relax IEEE, 2013, pp. 65--74.

\bibitem{Iannone2023Secret}
E.~Iannone, R.~Guadagni, F.~Ferrucci, A.~De~Lucia, and F.~Palomba, ``\BIBforeignlanguage{en}{The {{Secret Life}} of {{Software Vulnerabilities}}: {{A Large-Scale Empirical Study}}},'' \emph{\BIBforeignlanguage{en}{IEEE Transactions on Software Engineering}}, vol.~49, no.~1, pp. 44--63, 2023.

\bibitem{ExploitDB}
``Exploit database - exploits for penetration testers, researchers, and ethical hackers,'' \url{https://www.exploit-db.com/}, 2024, accessed: 2024-12-31.

\bibitem{xia2015learning}
P.~Xia, L.~Zhang, and F.~Li, ``Learning similarity with cosine similarity ensemble,'' \emph{Information sciences}, vol. 307, pp. 39--52, 2015.

\bibitem{agresti1998approximate}
A.~Agresti and B.~A. Coull, ``\BIBforeignlanguage{English}{Approximate is better than ``exact'' for interval estimation of binomial proportions},'' \emph{\BIBforeignlanguage{English}{The American Statistician}}, vol.~52, no.~2, pp. pp. 119--126, 1998.

\bibitem{li2024assessing}
S.~Li, Y.~Cheng, J.~Chen, J.~Xuan, S.~He, and W.~Shang, ``Assessing the performance of ai-generated code: A case study on github copilot,'' in \emph{2024 IEEE 35th International Symposium on Software Reliability Engineering (ISSRE)}.\hskip 1em plus 0.5em minus 0.4em\relax IEEE, 2024, pp. 216--227.

\bibitem{DBLP:conf/icse/DingCS20}
Z.~Ding, J.~Chen, and W.~Shang, ``Towards the use of the readily available tests from the release pipeline as performance tests: are we there yet?'' in \emph{{ICSE} '20: 42nd International Conference on Software Engineering, Seoul, South Korea, 27 June - 19 July, 2020}.\hskip 1em plus 0.5em minus 0.4em\relax {ACM}, 2020, pp. 1435--1446.

\bibitem{GitHub-github_api-11-12}
{GitHub, Inc.}, ``Github rest api,'' \url{https://docs.github.com/en/rest?apiVersion=2022-11-28}, 2024, accessed: 2024-11-12.

\bibitem{Tang2024Just}
\BIBentryALTinterwordspacing
X.~Tang, Z.~Chen, K.~Kim, H.~Tian, S.~Ezzini, and J.~Klein. Just-in-{{Time Detection}} of {{Silent Security Patches}}. [Online]. Available: \url{http://arxiv.org/abs/2312.01241}
\BIBentrySTDinterwordspacing

\bibitem{HuggingFace}
``Hugging face - the ai community building the future.'' \url{https://huggingface.co/}, 2025, accessed: 2025-02-18.

\bibitem{salton1983introduction}
G.~Salton and M.~J. McGill, \emph{Introduction to Modern Information Retrieval}.\hskip 1em plus 0.5em minus 0.4em\relax USA: McGraw-Hill, Inc., 1986.

\bibitem{sliwerski2005changes}
J.~{\'S}liwerski, T.~Zimmermann, and A.~Zeller, ``When do changes induce fixes?'' \emph{ACM sigsoft software engineering notes}, vol.~30, no.~4, pp. 1--5, 2005.

\bibitem{daniel2018biostatistics}
W.~W. Daniel and C.~L. Cross, \emph{Biostatistics: a foundation for analysis in the health sciences}.\hskip 1em plus 0.5em minus 0.4em\relax Wiley, 2018.

\bibitem{mchugh2012interrater}
M.~L. McHugh, ``Interrater reliability: the kappa statistic,'' \emph{Biochemia medica}, vol.~22, no.~3, pp. 276--282, 2012.

\bibitem{shi2022does}
J.~Shi, D.~Zou, S.~Xu, X.~Deng, and H.~Jin, ``Does openbsd and firefox’s security improve with time?'' \emph{IEEE Transactions on Dependable and Secure Computing}, vol.~20, no.~4, pp. 2781--2793, 2022.

\bibitem{SuperData:-2025-02-28}
U.~Joe~Durbin, ``Superdata: Vr's breakout 2016 saw 6.3 million headsets shipped | venturebeat,'' \url{https://venturebeat.com/business/superdata-vrs-breakout-2016-saw-6-3-million-headsets-shipped/}, 2025, accessed: 2025-02-28.

\bibitem{VR-2025-02-28}
``Vr headset shipments 'to boom' in 2016 - bbc news,'' \url{https://www.bbc.com/news/technology-36110422}, 2025, accessed: 2025-02-28.

\bibitem{Martin2017Virtual}
J.~Martín-Gutiérrez, C.~E. Mora, B.~Añorbe-Díaz, and A.~González-Marrero, ``\BIBforeignlanguage{en}{Virtual {{Technologies Trends}} in {{Education}}},'' \emph{\BIBforeignlanguage{en}{EURASIA Journal of Mathematics, Science and Technology Education}}, vol.~13, no.~2, 2017.

\bibitem{hall2017augmented}
S.~Hall and R.~Takahashi, ``Augmented and virtual reality: The promise and peril of immersive technologies,'' in \emph{World Economic Forum}, vol.~2, 2017.

\bibitem{Abbasi2017Technology}
M.~Abbasi, P.~Vassilopoulou, and L.~Stergioulas, ``\BIBforeignlanguage{en}{Technology roadmap for the {{Creative Industries}}},'' \emph{\BIBforeignlanguage{en}{Creative Industries Journal}}, vol.~10, no.~1, pp. 40--58, 2017.

\bibitem{chen2025unveiling}
H.~Chen, Z.~Huang, Y.~Xu, W.~Huang, J.~Chen, H.~Li, K.~Peng, F.~Liu, and S.~He, ``Unveiling code clone patterns in open source vr software: An empirical study,'' \emph{arXiv preprint arXiv:2501.07165}, 2025.

\bibitem{ernst2003static}
M.~D. Ernst, ``Static and dynamic analysis: Synergy and duality,'' in \emph{WODA 2003: ICSE Workshop on Dynamic Analysis}, 2003, pp. 24--27.

\bibitem{Altinkemer2008Vulnerabilitiesa}
K.~Altinkemer, J.~Rees, and S.~Sridhar, ``Vulnerabilities and patches of open source software: {{An}} empirical study,'' \emph{Journal of Information System Security}, vol.~4, no.~2, pp. 3--25, 2008.

\bibitem{Bosu2014Identifyinga}
A.~Bosu, J.~C. Carver, M.~Hafiz, P.~Hilley, and D.~Janni, ``\BIBforeignlanguage{en}{Identifying the characteristics of vulnerable code changes: An empirical study},'' in \emph{\BIBforeignlanguage{en}{Proceedings of the 22nd {{ACM SIGSOFT International Symposium}} on {{Foundations}} of {{Software Engineering}}}}.\hskip 1em plus 0.5em minus 0.4em\relax Hong Kong China: ACM, 2014, pp. 257--268.

\bibitem{DiPenta2009lifea}
M.~Di~Penta, L.~Cerulo, and L.~Aversano, ``The life and death of statically detected vulnerabilities: {{An}} empirical study,'' \emph{Information and Software Technology}, vol.~51, no.~10, pp. 1469--1484, 2009.

\bibitem{Zhang2011Empiricala}
S.~Zhang, D.~Caragea, and X.~Ou, ``An {{Empirical Study}} on {{Using}} the {{National Vulnerability Database}} to {{Predict Software Vulnerabilities}},'' in \emph{Database and {{Expert Systems Applications}}}, A.~Hameurlain, S.~W. Liddle, K.-D. Schewe, and X.~Zhou, Eds.\hskip 1em plus 0.5em minus 0.4em\relax Berlin, Heidelberg: Springer Berlin Heidelberg, 2011, vol. 6860, pp. 217--231.

\bibitem{Tufano2017When}
M.~Tufano, F.~Palomba, G.~Bavota, R.~Oliveto, M.~D. Penta, A.~De~Lucia, and D.~Poshyvanyk, ``When and {{Why Your Code Starts}} to {{Smell Bad}} (and {{Whether}} the {{Smells Go Away}}),'' \emph{IEEE Transactions on Software Engineering}, vol.~43, no.~11, pp. 1063--1088, 2017.

\bibitem{Li2017Large-Scale}
F.~Li and V.~Paxson, ``\BIBforeignlanguage{en}{A {{Large-Scale Empirical Study}} of {{Security Patches}}},'' in \emph{\BIBforeignlanguage{en}{Proceedings of the 2017 {{ACM SIGSAC Conference}} on {{Computer}} and {{Communications Security}}}}, 2017, pp. 2201--2215.

\bibitem{Shahzad2020Large}
M.~Shahzad, M.~Z. Shafiq, and A.~X. Liu, ``\BIBforeignlanguage{en}{Large {{Scale Characterization}} of {{Software Vulnerability Life Cycles}}},'' \emph{\BIBforeignlanguage{en}{IEEE Transactions on Dependable and Secure Computing}}, vol.~17, no.~4, pp. 730--744, 2020.

\bibitem{Liu2020large-scalea}
B.~Liu, G.~Meng, W.~Zou, Q.~Gong, F.~Li, M.~Lin, D.~Sun, W.~Huo, and C.~Zhang, ``\BIBforeignlanguage{en}{A large-scale empirical study on vulnerability distribution within projects and the lessons learned},'' in \emph{\BIBforeignlanguage{en}{Proceedings of the {{ACM}}/{{IEEE}} 42nd {{International Conference}} on {{Software Engineering}}}}.\hskip 1em plus 0.5em minus 0.4em\relax Seoul South Korea: ACM, 2020, pp. 1547--1559.

\bibitem{ShiDoes}
J.~Shi, D.~Zou, S.~Xu, X.~Deng, and H.~Jin, ``Does openbsd and firefoxâs security improve with time?'' \emph{IEEE Transactions on Dependable and Secure Computing}, vol.~20, no.~4, pp. 2781--2793, 2022.

\bibitem{Oyelere2020Exploringa}
S.~S. Oyelere, N.~Bouali, R.~Kaliisa, G.~Obaido, A.~A. Yunusa, and E.~R. Jimoh, ``\BIBforeignlanguage{en}{Exploring the trends of educational virtual reality games: A systematic review of empirical studies},'' \emph{\BIBforeignlanguage{en}{Smart Learning Environments}}, vol.~7, no.~1, p.~31, 2020.

\bibitem{Grudzewski2018Virtuala}
G.~Filip, A.~Marcin, M.~Grzegorz, and P.~Katarzyna, ``{Virtual Reality in Marketing Communication – the Impact on the Message, Technology and Offer Perception – Empirical Study},'' \emph{Economics and Business Review}, vol.~4, no.~3, pp. 36--50, July 2018.

\bibitem{Epp2021empirical}
R.~Epp, D.~Lin, and C.-P. Bezemer, ``An empirical study of trends of popular virtual reality games and their complaints,'' \emph{IEEE Transactions on Games}, vol.~13, no.~3, pp. 275--286, 2021.

\bibitem{Huang2024Study}
W.~Huang, J.~Chen, H.~Chen, Z.~Qi, X.~Yang, K.~Peng, and S.~He, ``A study of code clone on open source vr software,'' in \emph{Proceedings of the 39th IEEE/ACM International Conference on Automated Software Engineering Workshops}, 2024, p. 239–244.

\bibitem{Lenarduzzi2020OpenSZZ}
V.~Lenarduzzi, F.~Palomba, D.~Taibi, and D.~A. Tamburri, ``\BIBforeignlanguage{en}{{{OpenSZZ}}: {{A Free}}, {{Open-Source}}, {{Web-Accessible Implementation}} of the {{SZZ Algorithm}}},'' in \emph{\BIBforeignlanguage{en}{Proceedings of the 28th {{International Conference}} on {{Program Comprehension}}}}.\hskip 1em plus 0.5em minus 0.4em\relax ACM, 2020, pp. 446--450.

\bibitem{Borg2019SZZ}
M.~Borg, O.~Svensson, K.~Berg, and D.~Hansson, ``\BIBforeignlanguage{en}{{{SZZ Unleashed}}: {{An Open Implementation}} of the {{SZZ Algorithm}} -- {{Featuring Example Usage}} in a {{Study}} of {{Just-in-Time Bug Prediction}} for the {{Jenkins Project}}},'' in \emph{\BIBforeignlanguage{en}{Proceedings of the 3rd {{ACM SIGSOFT International Workshop}} on {{Machine Learning Techniques}} for {{Software Quality Evaluation}}}}, 2019, pp. 7--12.

\bibitem{Jira}
Atlassian, ``Jira | issue \& project tracking software | atlassian,'' \url{https://www.atlassian.com/software/jira}, 2024, accessed: 2024-12-31.

\bibitem{Bugzilla}
``Bugzilla,'' \url{https://www.bugzilla.org/}, 2024, accessed: 2024-12-31.

\bibitem{mikolov2013efficient}
T.~Mikolov, ``Efficient estimation of word representations in vector space,'' \emph{arXiv preprint arXiv:1301.3781}, vol. 3781, 2013.

\end{thebibliography}



\vspace{-1cm}

\begin{IEEEbiography}
[{\includegraphics[width=1in,height=1.25in,clip,keepaspectratio]{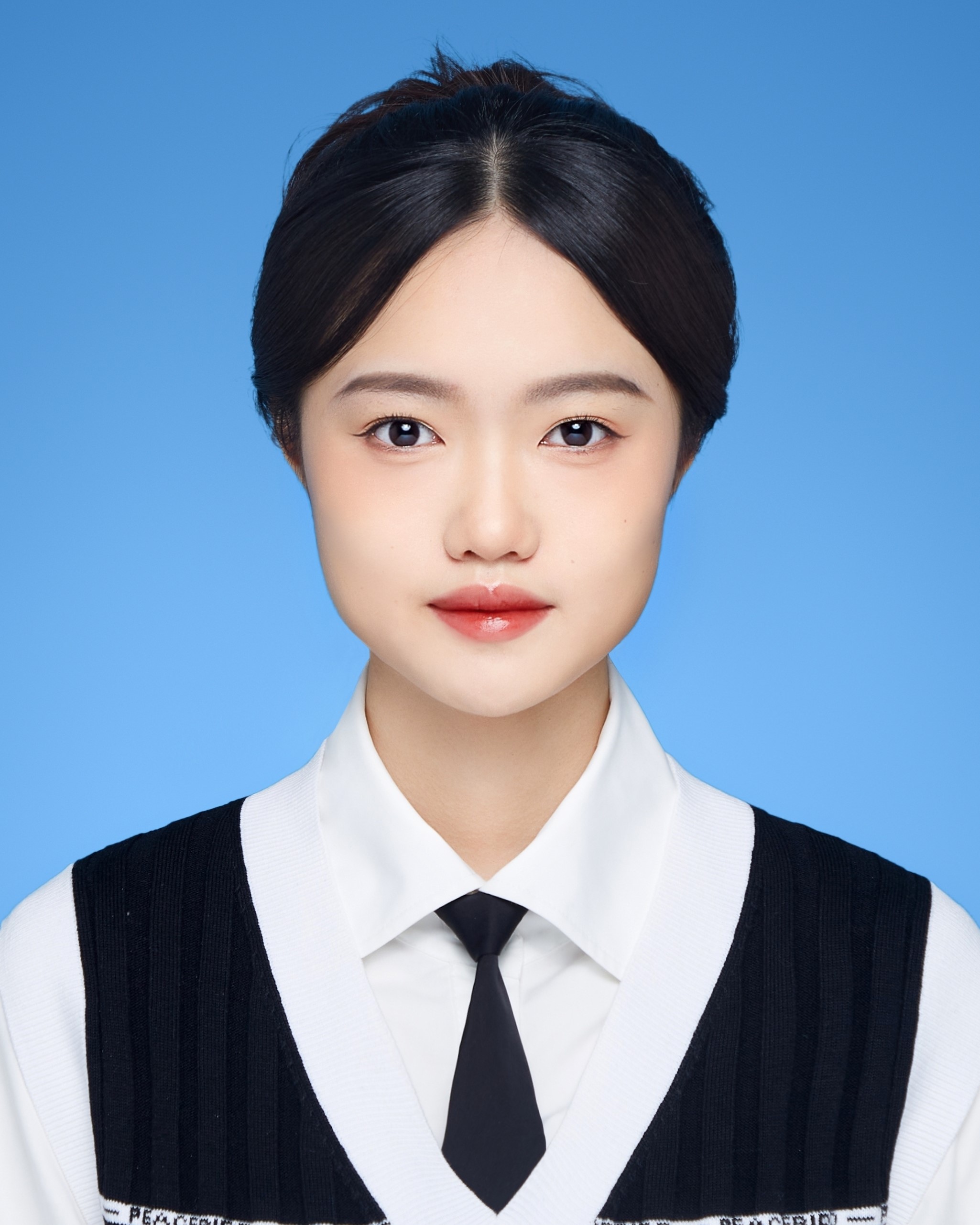}}]
{Yifan Xu} received the BSc degree from Software College of Northeastern University, Liaoning, China, in 2023. She is currently working toward the Ph.D. degree with Institute of Information Engineering, Chinese Academy of Sciences, Beijing, China. Her research of interests are in the areas of cyberspace security and biometric authentication.
\end{IEEEbiography}

\vspace{-1.5cm}

\begin{IEEEbiography}
[{\includegraphics[width=1in,height=1.25in,clip,keepaspectratio]{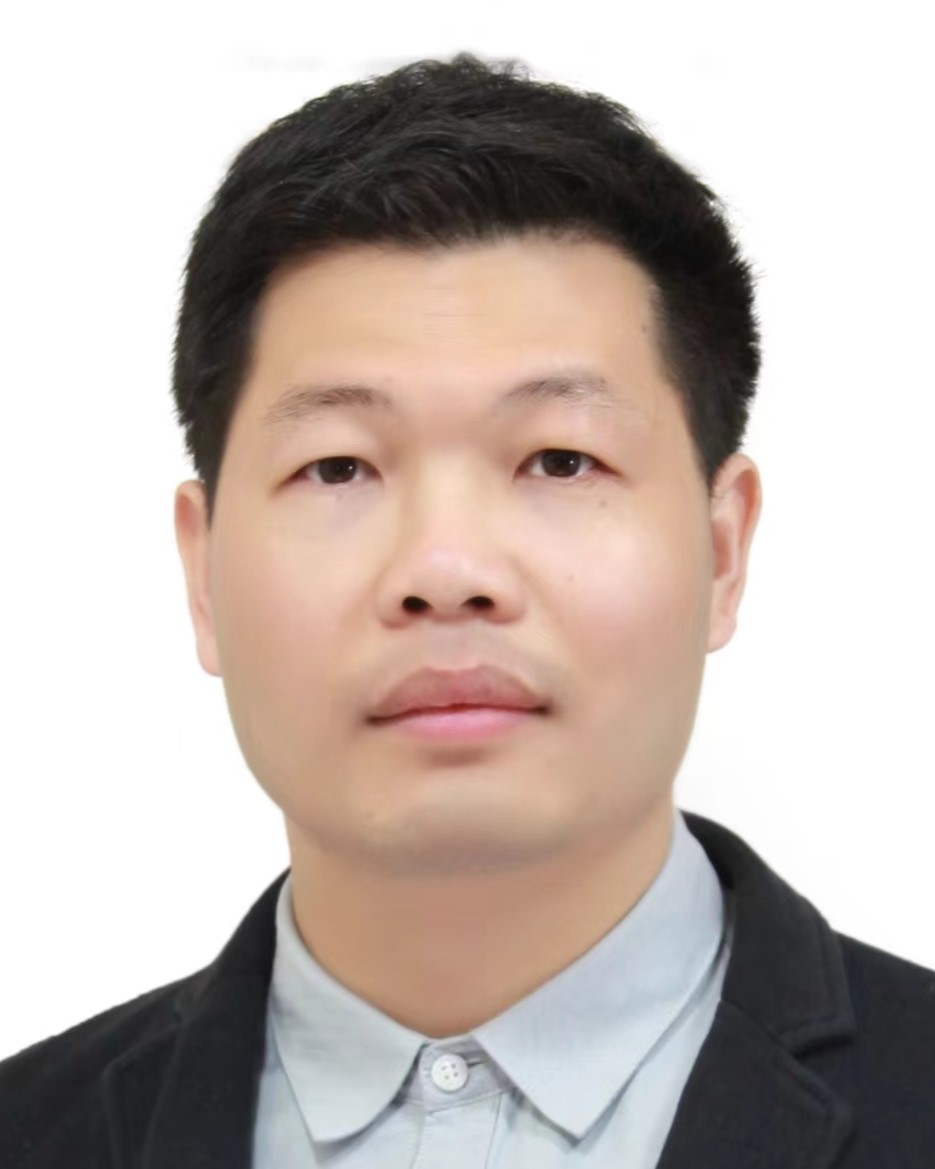}}]
{Jinfu Chen} is an Associate Professor at School of Computer Science, Wuhan University. His research interest lies in software performance engineering, software performance testing, software log mining, and code clone detection. His work has been published in flagship conferences and journals such as ICSE, FSE, ASE, TSE, TOSEM, and TIFS. He is a reviewer for software engineering journals such as TSE, JSS, EMSE. Dr Jinfu Chen got his Bachelor’s degree at Harbin Institute of Technology, a Master’s degree from the Chinese Academy of Sciences, and received a Ph.D. at Concordia University.
\end{IEEEbiography}

\vspace{-1cm}

\begin{IEEEbiography}
[{\includegraphics[width=1in,height=1.25in,clip,keepaspectratio]{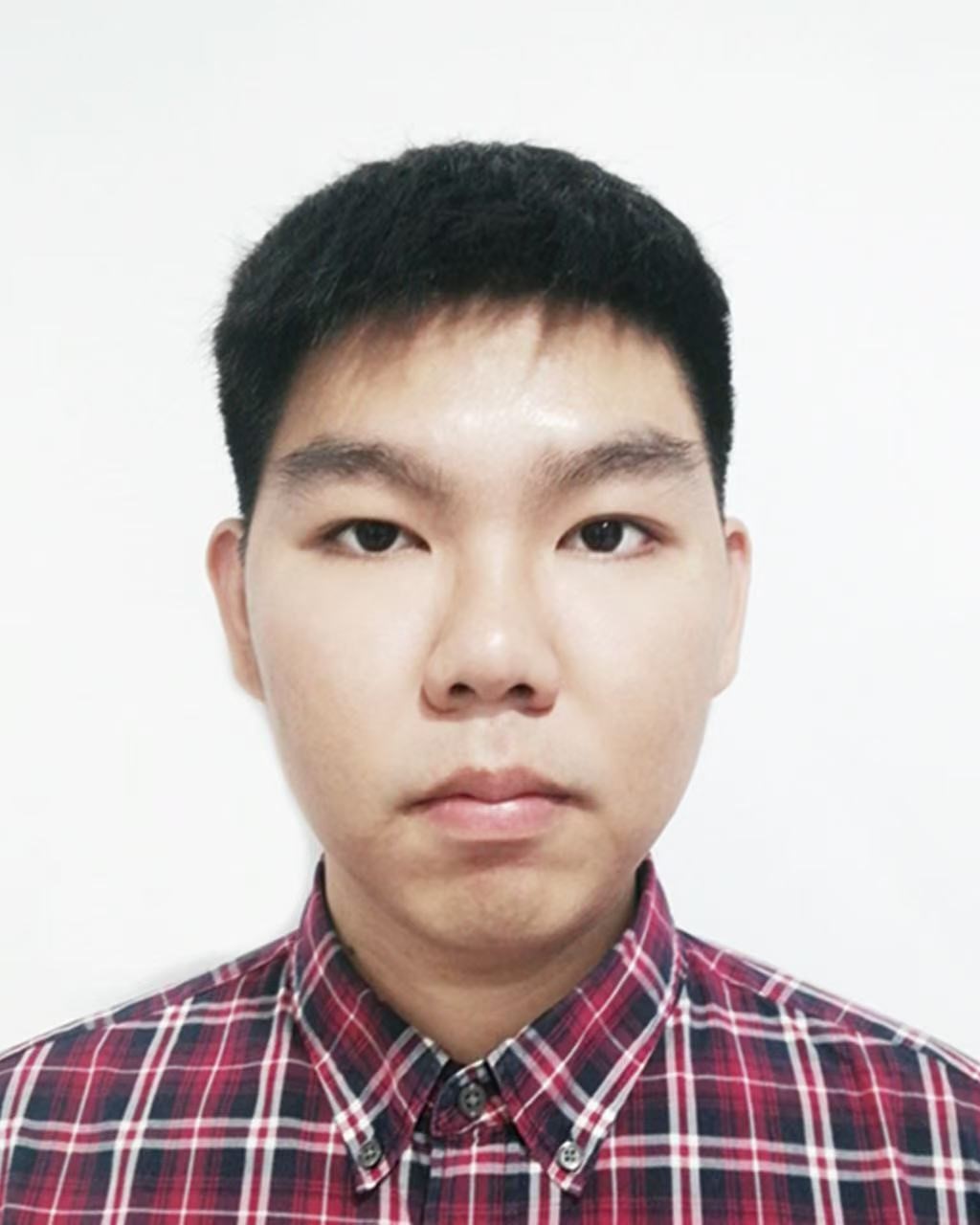}}]
{Zhenyu Qi} is a Ph.D. student in the Department of Electrical and Computer Engineering at the University of Arizona, advised by Dr. Sen He. His research interests lie in AI for Software Engineering (AI4SE) and Quantum Computing. Specifically, his existing work focuses on leveraging LLMs for software development automation and LLM for Quantum Computing. Zhenyu has published in the ASE conference, and AUSE Journal.
\end{IEEEbiography}

\vspace*{-2\baselineskip}

\begin{IEEEbiography}
[{\includegraphics[width=1in,height=1.25in,clip,keepaspectratio]{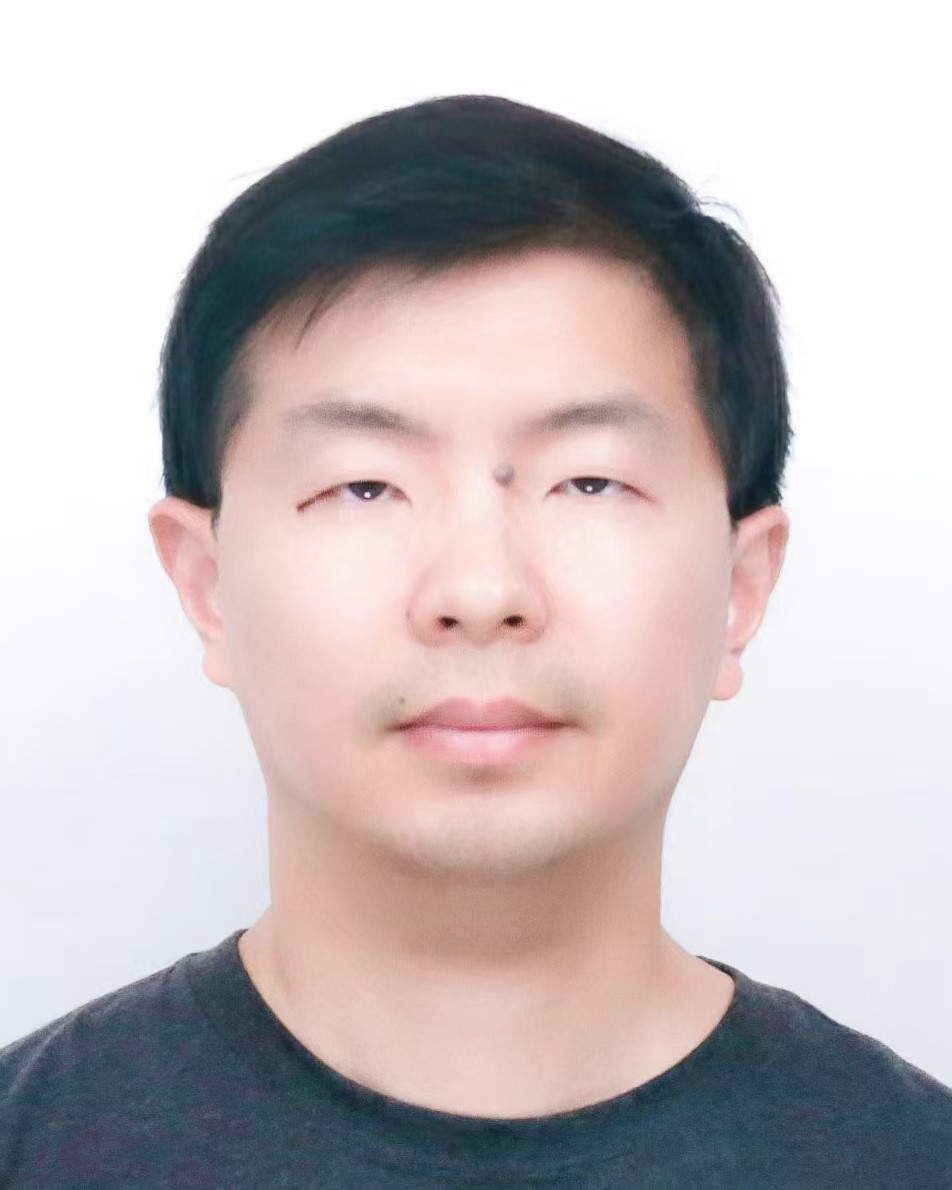}}]
{Huashan Chen} is an Associate Professor at the Institute of Information Engineering, Chinese Academy of Sciences. He received the B.S. and M.S. degrees from Shandong University, and the Institute of Information Engineering, Chinese Academy of Sciences, respectively. He received the Ph.D. degree in Computer Science from The University of Texas at San Antonio. His work has been published in premier conferences and journals including CSUR, TDSC, TIFS, TKDE, KDD and SIGIR. His research interests lie in VR/AR Security \& Privacy.
\end{IEEEbiography}

\vspace{-0.8cm}

\begin{IEEEbiography}
[{\includegraphics[width=1in,height=1.25in,clip,keepaspectratio]{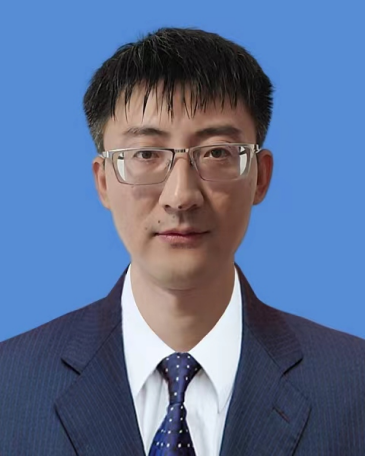}}]
{Junyi Wang} is an Assistant Professor at School of Computer Science and Technology, Shandong University. He obtained his Ph.D. at State Key Laboratory of Virtual Reality Technology and Systems, Beihang University. His research covers Virtual Reality/Augmented Reality, camera localization, object pose estimation and 3D reconstruction.
\end{IEEEbiography}

\vspace{-0.8cm}

\begin{IEEEbiography}[{\includegraphics[width=1in,height=1.25in,clip,keepaspectratio]{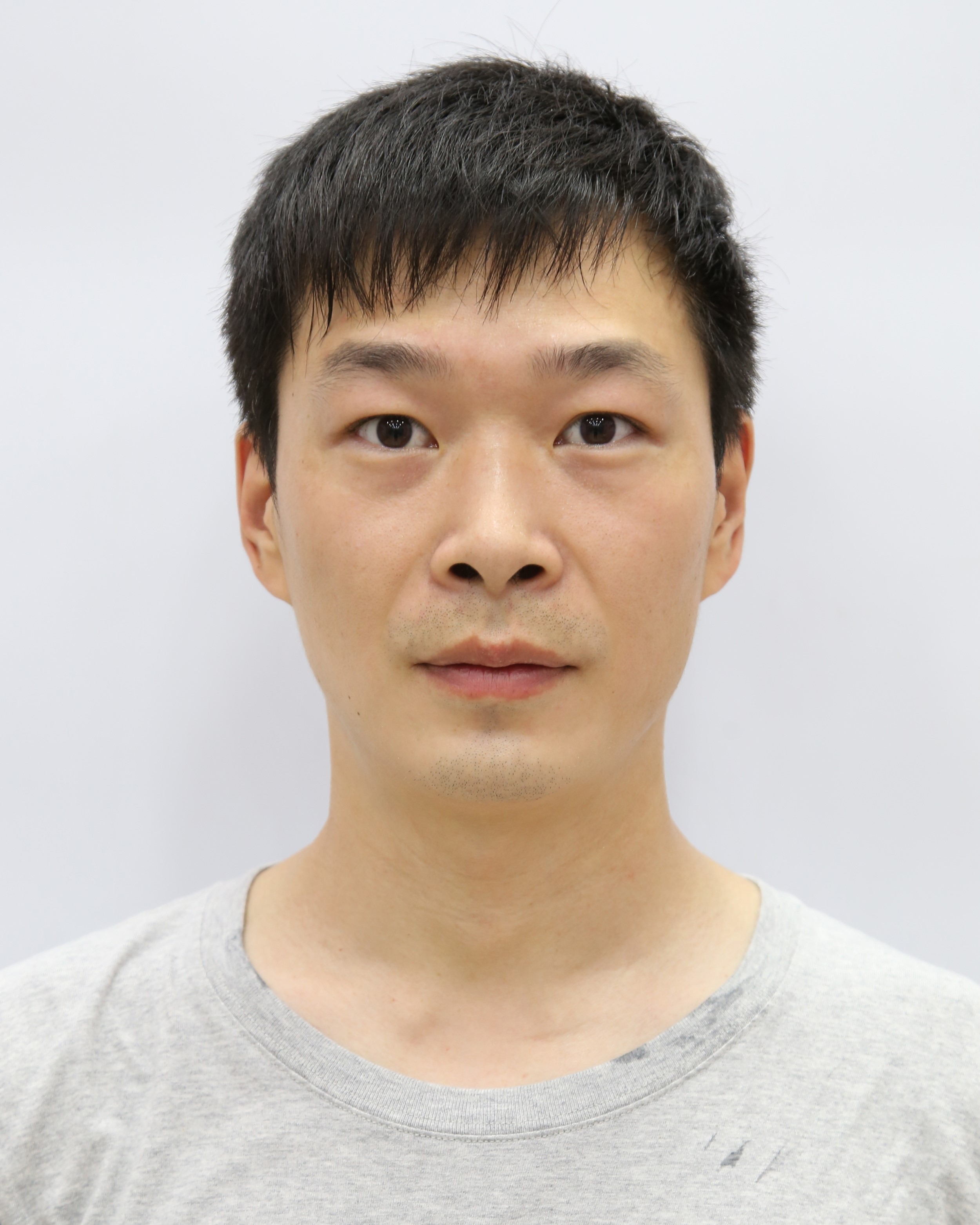}}]
{Pengfei Hu} is a Professor in the School of Computer Science and Technology at Shandong University. He received Ph.D. in Computer Science from UC Davis. His research interests are in the areas of cyber security, data privacy, and mobile computing. He has published more than 60 papers in premier conferences and journals on these topics. He served as TPC for numerous prestigious conferences, and associate editors for IEEE TDSC, IEEE TWC and IEEE IoTJ.
\end{IEEEbiography}

\vspace{-0.8cm}

\begin{IEEEbiography}
[{\includegraphics[width=1in,height=1.25in,clip,keepaspectratio]{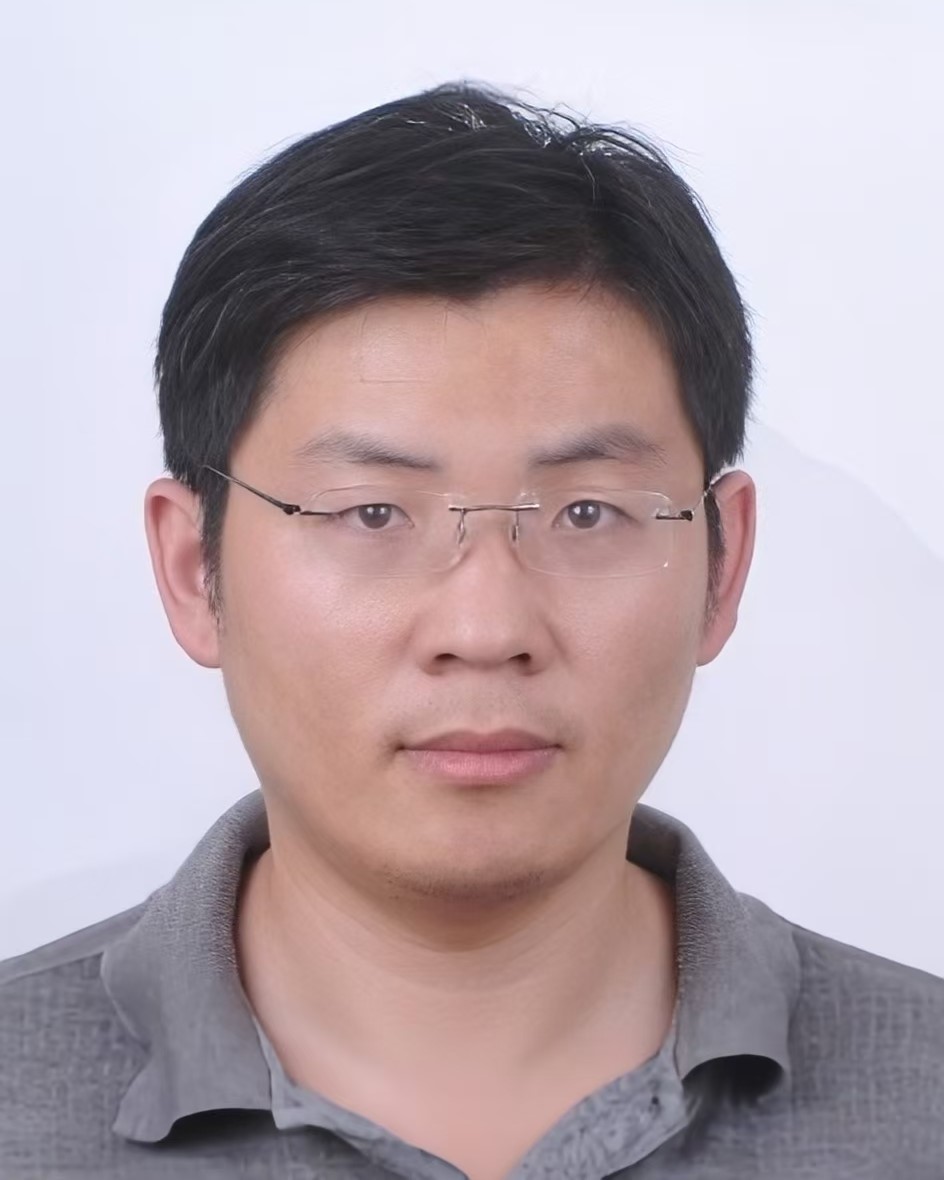}}]
{Feng Liu} is a Professor and Ph.D. supervisor at the Institute of Information Engineering, Chinese Academy of Sciences. He is also a professor in the School of Cyber Security, University of Chinese Academy of Sciences. He received his B.S. degree in 2003 from Shandong University and the Ph.D. degree in 2009 from Institute of Software, Chinese Academy of Sciences. His research interests include system security, visual security and cryptography. He co-initiated the International Conference on Science of Cyber Security (SciSec) and is serving as its Steering Committee Chair. He serves as the Editor-in-Chief of the International Journal of Digital Crime and Forensics (IJDCF).  
\end{IEEEbiography}

\vspace{-0.8cm}

\begin{IEEEbiography}
[{\includegraphics[width=1in,height=1.25in,clip,keepaspectratio]{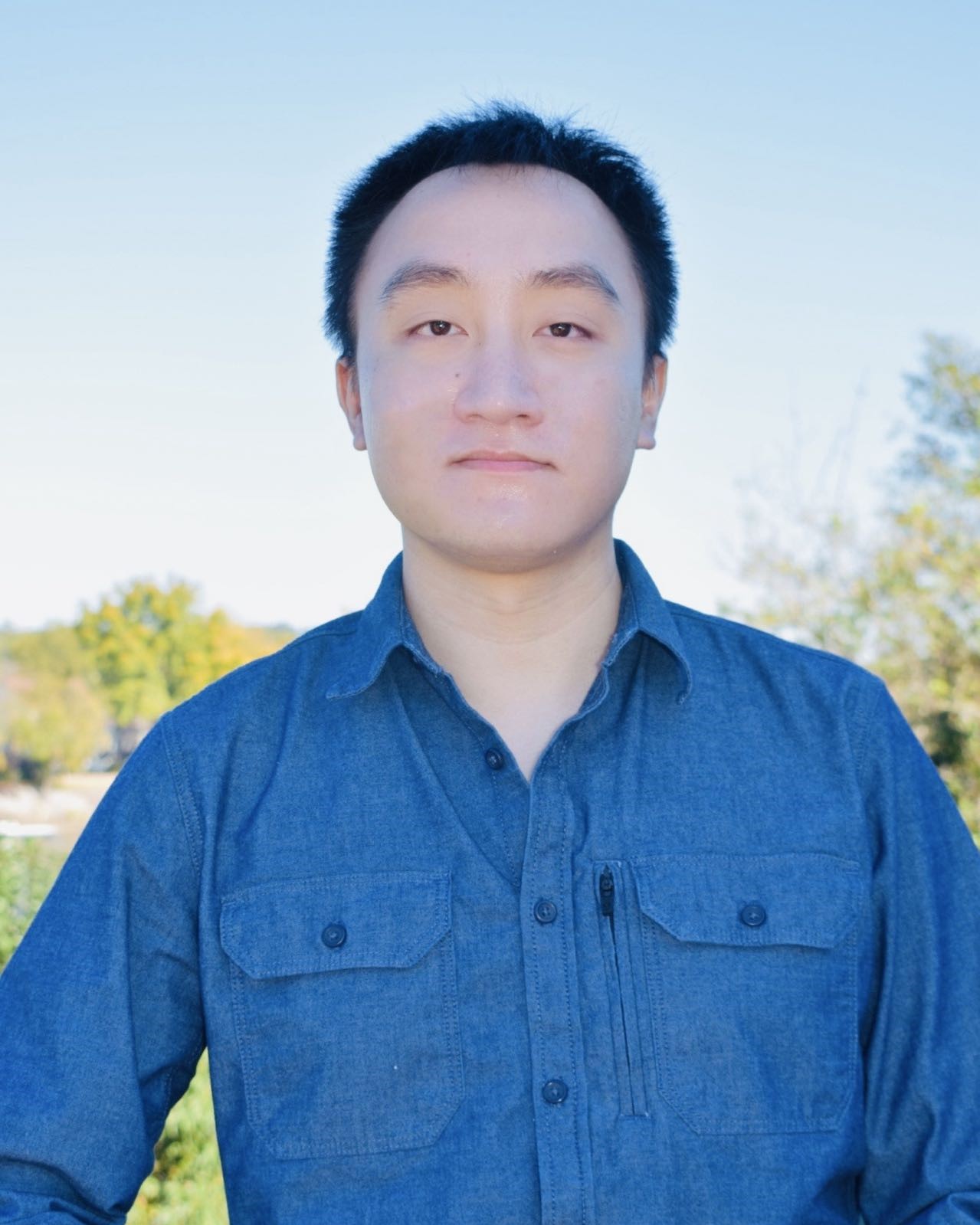}}]
{Sen He} is an Assistant Professor in the Department of Electrical and Computer Engineering at the University of Arizona. Sen He received his Ph.D. from The University of Texas at San Antonio in July 2022. Sen He’s research interests include the areas of Cloud Computing, Software/Performance Engineering (Formal Methods, LLM for SE), and Computer Vision. His works have been published in flagship conferences, including ESEC/FSE, MICRO, ASE, ECCV, and EuroSys. And one of his main works has received the ACM SIGSOFT Distinguished Paper Award.
\end{IEEEbiography}

\vfill

\newpage
\appendices 
\section{Process of Combined Traditional Method: TF-IDF~and~Wor2Vec}
\label{sec:appendix_tfidf}

To classify security-related commits under the CWE-699 category, we propose a text-based approach that combines word weighting, sentence vector construction, and similarity-based classification.

\subsection{Word Weighting}
For each weakness category under CWE-699, we first preprocess the associated descriptions by removing standard stop words and phrases that lack semantic meaning. Then, we apply the TF-IDF method to determine the importance of each word within the descriptions. Specifically, we calculate the frequency of word occurrence in each weakness description, known as Term Frequency (TF). The TF for a given word \(word_i\) in category \(cat_j\) is defined as:
\begin{equation}
    TF_{i,j} = \frac{n_{i,j}}{{\textstyle \sum_{k=1}^{40}}n_{k,j}}
\end{equation}
where \(n_{i,j}\) denotes the number of times \(word_i\) appears in the description of category \(cat_j\). After getting TF, we calculate the proportion of categories containing each word, known as Inverse Document Frequency (IDF). The IDF for \(word_i\) is defined as:
\begin{equation}
    IDF_i=\lg{\frac{40}{1 + \left | j:word_i\in cat_j \right | }} 
\end{equation}
where \(\left | j:word_i\in cat_j \right |\) represents the number of categories that contain \(word_i\) in weakness descriptions. After obtaining TF and IDF, we proceed to compute the importance score for each word. Formally, the important score of \(word_i\) is defined as:
\begin{equation}
    TFIDF_{i,j}=TF_{i,j} \times IDF_i
    \label{eq:tf-idf}
\end{equation}
In this way, we determine the word weight for each of the 40 weakness categories within CWE-699.

For each security-related commit message, we apply the same processing method and calculate the weight of \(word_i\) using the following formula:
\begin{equation}
    TFIDF'_{i,sen}= \frac{n_{i,sen}}{{\textstyle \sum_{k=1}^{40}}n_{k,j}} \times IDF_{i}
    \label{eq:commit_tf-idf}
\end{equation}
where \(sen\) refers to the commit message, and \(n_{i,sen}\) represents how many times \(word_i\) appears in the message \(sen\).

\subsection{Sentence Vector Construction}
We construct sentence vectors by applying word weights to the weakness descriptions of the 40 categories within CWE-699 and to security-related commit messages.

For each weakness category under CWE-699, we first generate an initial word vector for each word in the sentence using a pre-trained Word2Vec model \cite{mikolov2013efficient}, which produces a semantic representation of the word in a high-dimensional vector space. 
Next, each word vector is multiplied by its corresponding weight derived from (\ref{eq:tf-idf}). Finally, we aggregate all the word vectors from the descriptions to form the sentence vector. This process can be formally represented as follows:
\begin{equation}
        V_{weighted}^j = \frac{{\textstyle \sum_{i=1}^N}TFIDF_{i,j} \cdot v_i^j}{{\textstyle \sum_{i=1}^N}TFIDF_{i,j}}
\end{equation}
where \(V_{weighted}^j\) represents the weighted sentence vector for category \(cat_j\),  \(N\) is the number of words in the description of the category, \(v_i^j\) denotes the word vector of \(word_i\) produced by Word2Vec, and \(TFIDF_{i,j}\) is defined in (\ref{eq:tf-idf}).

For each security-related commit message, we follow the same procedure outlined above. The formula for forming the sentence vectors of a commit message \(sen\) is as follows:
\begin{equation}
        {V'}_{weighted}^{sen} = \frac{{\textstyle \sum_{i=1}^N}TFIDF'_{i,sen} \cdot v_i^{sen}}{{\textstyle \sum_{i=1}^N}TFIDF'_{i,sen}}
\end{equation}
where \(TFIDF_{i,sen}\) is defined in (\ref{eq:commit_tf-idf}).

\subsection{Similarity-based Classification}
After constructing the sentence vectors for both CWE-699 weakness descriptions and commit messages, we calculate the cosine similarity \cite{xia2015learning} between an input commit message and each weakness description within CWE-699 for two main purposes: (i) to assess whether the commit pertains to resolving an existing security weakness (i.e., to determine if the commit is a WFC). If the cosine similarity between the commit message and any of the 40 weakness descriptions exceeds zero, the commit is classified as a WFC; (ii) to classify the identified WFC into one of the 40 categories within CWE-699. The category with the highest cosine similarity score is considered the most relevant category to the commit message.

\end{document}